# An All-Atom Model of the Resorcinol–Formaldehyde Aerogel Skeleton from DFT and Molecular Dynamics: Validation and Iron Incorporation

Shoeb ATHAR[1,2], Jérémy GUAZZAGALOPPA[1,2], and Philippe JUND[1,*]

[1] ICGM, Univ Montpellier, CNRS, ENSCM, 34293 Montpellier, France

[2] HUTCHINSON SA - Rue Gustave Nourry, 45120 Châlette-sur-Loing, France

* Corresponding author. E-mail address: philippe.jund@umontpellier.fr (Philippe JUND)

## Abstract

Resorcinol–formaldehyde (RF) aerogels are lightweight, ultra-porous thermal superinsulators whose complex amorphous structure has not yet been described at the all-atom level. Here, we build an all-atom model of the RF aerogel skeleton by combining density functional theory (DFT) and classical molecular dynamics (MD), validate it against experimental infrared and density measurements, and use it to study the incorporation of iron dopants. The RF base unit was optimized with Gaussian, $DMol^3$ and SIESTA, and its calculated infrared spectrum agreed with experiment. Comparing classical force fields with DFT for branched systems of two to four base units (140–278 atoms) led to the choice of COMPASS II. A 15,012-atom system of 54 branched RF units, not cross-linked to each other, aggregated into a single cluster in the canonical (*NVT*) ensemble and reached an equilibrium density of 1.296 g $cm^{-3}$ in the isobaric–isothermal (*NPT*) ensemble, in agreement with the experimental skeletal density (1.3 g $cm^{-3}$). Then, 5 wt% iron (109 atoms) was added to the equilibrated structure: during 1 ns of *NPT* simulation, the fraction of iron atoms embedded in the skeleton rose from 63% to 71%, and the Fe–C radial distribution function peaked at 3 Å. The model thus provides an experimentally supported starting point for screening fillers and dopants in RF gels, with prospective application to their thermoelectric functionalization, which we have previously demonstrated experimentally with basalt and PANOX fibres. Future extensions could address skeleton properties relevant to thermal insulation and the pyrolysis of RF gels into carbon and metal-doped carbon aerogels.

Keywords: Resorcinol–formaldehyde aerogel; Molecular dynamics; Density functional theory; COMPASS II force field; Iron doping

## 1. Introduction

Resorcinol–formaldehyde (RF) aerogels, first prepared by Pekala through the sol–gel polycondensation of resorcinol with formaldehyde [1], are lightweight, highly porous organic solids with a very low thermal conductivity, down to about 12 mW $m^{-1}$ $K^{-1}$ in air at room temperature for a density of about 157 kg $m^{-3}$ [2]. Their density and pore structure can be tuned through the synthesis parameters, and they are also widely used as precursors of carbon aerogels [3]. These properties make RF gels attractive as lightweight thermal superinsulators.

Thermoelectric (TE) materials convert a temperature difference directly into electrical energy through the Seebeck effect, and vice versa, which makes them attractive for waste-heat recovery and solid-state cooling. Their efficiency is measured by the dimensionless figure of merit $ZT = S^2\sigma T/\kappa$, where $S$ is the Seebeck coefficient, $\sigma$ the electrical conductivity, $\kappa$ the thermal conductivity and $T$ the absolute temperature. A good TE material must therefore combine a high power factor, $S^2\sigma$, with a low thermal conductivity, two requirements that are difficult to satisfy simultaneously [4]. Thermal superinsulators such as aerogels already meet the second requirement but are electrical insulators; their electrical functionalization, by doping or by the insertion of conductive fillers, is thus a possible route towards lightweight and inexpensive TE materials. Organic materials are well suited to this approach: electronically conducting organic materials, whose importance was recognized by the 2000 Nobel Prize in Chemistry awarded to A. MacDiarmid, A. Heeger and H. Shirakawa [5], can become semiconducting through their synthesis or doping while retaining the mechanical properties of polymers, owing in part to their $\pi$-conjugated carbon chains [6].

We have previously explored this route. For a silica-based superinsulating material, Aerosil 200, a numerical model built with density functional theory (DFT) and the BoltzTraP code gave a Seebeck coefficient, an electrical conductivity and a power factor in excellent agreement with experiment [7]. For RF gels, we showed that combining pyrolysis with the insertion of basalt or PANOX fibres increases the electrical conductivity and $ZT$ by twelve orders of magnitude, while the thermal conductivity remains below 50 mW $m^{-1}$ $K^{-1}$ [8]. Further improvement now depends on identifying the most effective fillers and dopants. Experimentally, this search is slow and costly, because of the large number of synthesis parameters and candidate dopants, and because the complex amorphous structure of these materials makes the mechanisms involved difficult to understand.

Materials development has traditionally relied on empirical optimization guided by chemical knowledge, an approach that becomes inefficient when the number of candidate compositions is very large. Predictive computational modelling provides a more rational route: first-principles calculations can predict the electronic properties of materials, reduce the number of compositions that need to be synthesized, and help to identify the mechanisms at the origin of these properties. For amorphous polymer networks such as RF gels, however, these calculations first require a realistic atomistic model of the structure. Atomistic molecular dynamics (MD) models of cross-linked phenolic (phenol–formaldehyde) resins have been developed for more than a decade [9–11], but the first MD study of resorcinol resins, based on a united-atom description and a geometry-based cross-linking algorithm, was published only in 2024 [12]. In a united-atom description, each carbon atom and the hydrogen atoms bonded to it are merged into a single interaction site, whereas an all-atom model represents every atom, including hydrogen, explicitly, with its own partial charge and force-field parameters, so that electrostatic interactions and hydrogen bonding are described directly. To our knowledge, the present work provides the first all-atom model of the RF aerogel skeleton validated against both spectroscopic and density data, and the first simulation of dopant incorporation in it.

The model is built step by step. The RF base unit is first optimized by DFT with three codes and compared with the experimental infrared spectrum; a classical force field is then selected by comparison with DFT for branched systems of up to four base units; a 15,012-atom system of these units is built, equilibrated by MD in the canonical and isobaric–isothermal ensembles and compared with the experimental skeletal density; finally, the incorporation of iron particles, chosen as a first dopant, is examined. Because the branched units are not covalently cross-linked to each other but aggregate through non-bonded interactions, the resulting structure is an aggregated all-atom representation of the RF skeleton rather than a fully connected RF network. The TE properties themselves are not calculated here; the model is intended as the starting point for such calculations. The model is also relevant beyond thermoelectrics, from thermal insulation to the carbon aerogels derived from RF gels.

## 2. Computational Methods

### 2.1 DFT Calculations

The geometry of the RF base unit, of composition $C_{30}H_{30}O_{11}$ (71 atoms), was optimized by density functional theory (DFT) calculations, and the larger systems were then treated by molecular dynamics, using the Materials Studio 2017 software developed by Dassault Systèmes [13]. Among the quantum mechanics modules it offers, we used Gaussian 09 [14] and $DMol^3$ [15, 16]; the SIESTA program was also used [17–19]. Gaussian, named after its use of Gaussian orbitals, makes it possible to predict the structure, the energies, the thermodynamic properties, the NMR spectrum and the vibrational spectra of molecular systems [20]. DMol$^3$ models the electronic and energetic structure of molecules, solids and surfaces by DFT with very accurate results at a fairly low computational cost for an *ab initio*-type method. SIESTA (Spanish Initiative for Electronic Simulations with Thousands of Atoms) owes its efficiency to strictly localized basis sets. The functionals and basis sets used are given in Table 1; all calculations on the base unit were performed with free boundary conditions.

### 2.2 Vibrational Frequencies and Infrared Spectrum

The harmonic vibrational frequencies and infrared intensities of the RF base unit were calculated with SIESTA for the SIESTA-optimized structure. The frequencies are obtained by expanding the potential energy in a Taylor series to second order around the equilibrium position and diagonalizing the mass-weighted Hessian matrix [21, 22]. The calculated infrared spectrum was compared with an experimental spectrum to validate the RF base unit (Section 3.1).

### 2.3 Classical Force Fields

Classical calculations were performed with the Forcite module, whose key approximation is that the potential energy surface on which the atomic nuclei move is represented by a classical force field, with parameters fitted to experimental data and to quantum mechanics calculations; Forcite operates with the Velocity Verlet algorithm. Four of its force fields were compared (Section 3.2): Dreiding [23], CVFF [24], PCFF [25, 26] and COMPASS II [27–30], the last of which is parameterized on *ab initio* data and optimized for condensed phases. These force fields are described in SI Section S1.1.

### 2.4 Molecular Dynamics Protocol

Newton's equations of motion were integrated with the Velocity Verlet algorithm [31], which is equivalent to the Verlet algorithm [32] and is very widely used in molecular dynamics codes; it has the advantage of being easily programmable and of great numerical stability, making simulation over very long times possible. The time step was chosen by a symplecticity test in the microcanonical (*NVE*) ensemble (Section 3.3). Since the *NVE* ensemble does not generally reflect experimental conditions at constant temperature and/or pressure, the simulations were carried out in the canonical (*NVT*) and isobaric–isothermal (*NPT*) ensembles, the system being coupled to a thermostat or to a barostat, respectively. We chose the Nosé–Hoover thermostat [33–37], which describes the interactions between the heat bath and the system while respecting the laws of thermodynamics, and the Berendsen barostat [38], which changes the size of the simulation box uniformly, as recommended for the study of amorphous materials. With the selected force field, geometry optimizations used the conjugate gradient method [39] and the long-range forces were calculated by Ewald summation [40].

The periodic starting structures were built with the Amorphous Cell module, which inserts molecules into a cell randomly by the Monte Carlo method, minimizing the contacts between atoms while ensuring a distribution of the torsion angles for a given force field (COMPASS II in our case), for a defined number of particles at a chosen density [41].

## 3. Results and Discussion

### 3.1 RF Base Unit: DFT Optimization and Infrared Validation

The first objective was to represent the association of the resorcinol and formaldehyde precursors in agreement with the literature [42]. The geometry of the RF base unit, of composition $C_{30}H_{30}O_{11}$, was therefore optimized with free boundary conditions using three software packages: Gaussian and DMol$^3$ with the B3LYP functional, and SIESTA with the GGA-PBE functional (Table 1).

**Table 1.** Calculation parameters and results relating to the geometry optimization of the RF base unit.

| **Software** | **Gaussian** | **DMol$^3$** | **SIESTA** |
|---|---|---|---|
| Functional | B3LYP | B3LYP | GGA-PBE |
| Basis set | 6-31G(d,p) | DNP | DZP |
| **Energy (eV/part)** | **−762.219** | **−762.204** | **−761.998** |
| Structures | 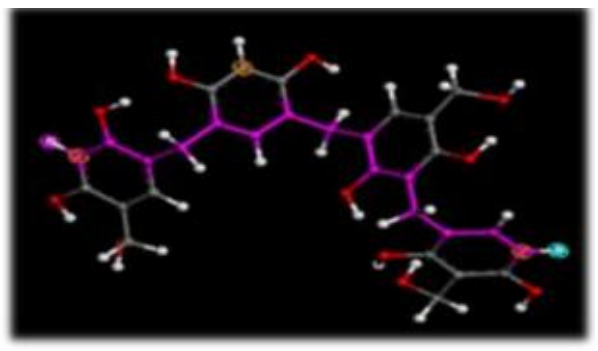 | 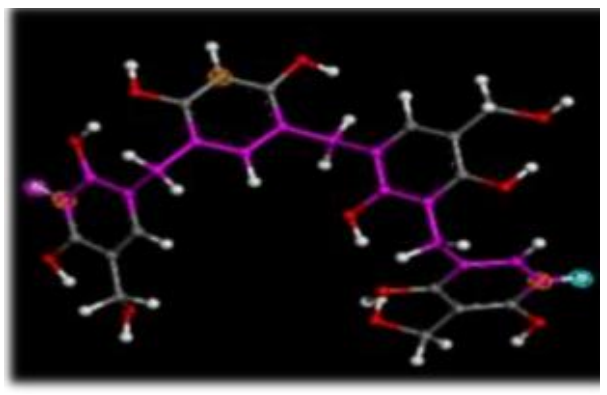 | 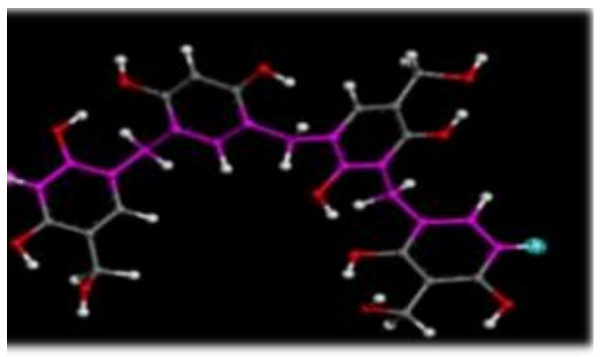 |

The geometry optimization results from the three software packages are similar, both from an energetic and from a structural point of view, which gives confidence in the structure obtained. To objectively compare their similarity, we calculated the pair distribution functions, $g(r)$, of the three structures. This function depends directly on the distribution of the interatomic distances around a chosen atom of the structure: it quantifies the modulations of the atomic density $\rho(r)$ relative to the mean atomic density $\rho_0$ as a function of the interatomic distance $r$, and has the advantage that it can be measured experimentally by diffraction [43]. Its definition is given in SI Section S1.2 (Eq. S1). The $g(r)$ of the three structures are compared in Figure 1. The first-neighbour peaks of the three structures lie within about 0.02 Å of each other; they differ mainly in height, the SIESTA O–H peak being lower and broader.

Upon closer inspection, the geometry optimization from SIESTA gives a structure slightly different from the other two, notably due to an inversion of the central carbon. However, this configuration is as stable as the others, since the interactions between nearest neighbors remain unchanged. In parallel, at each iteration of the SIESTA geometry optimization, we calculated the highest interaction forces on each atom of the RF molecule along the three directions of space, to ensure that no residual interaction force remained at the end of the modelling. Only about ten atoms concentrated the largest forces during the optimization (Figure 2), and at the end of the calculation no force greater than $10^{-4}$ eV Å$^{-1}$ remained in the system (Figure 3), which shows that the RF base unit has been correctly optimized.

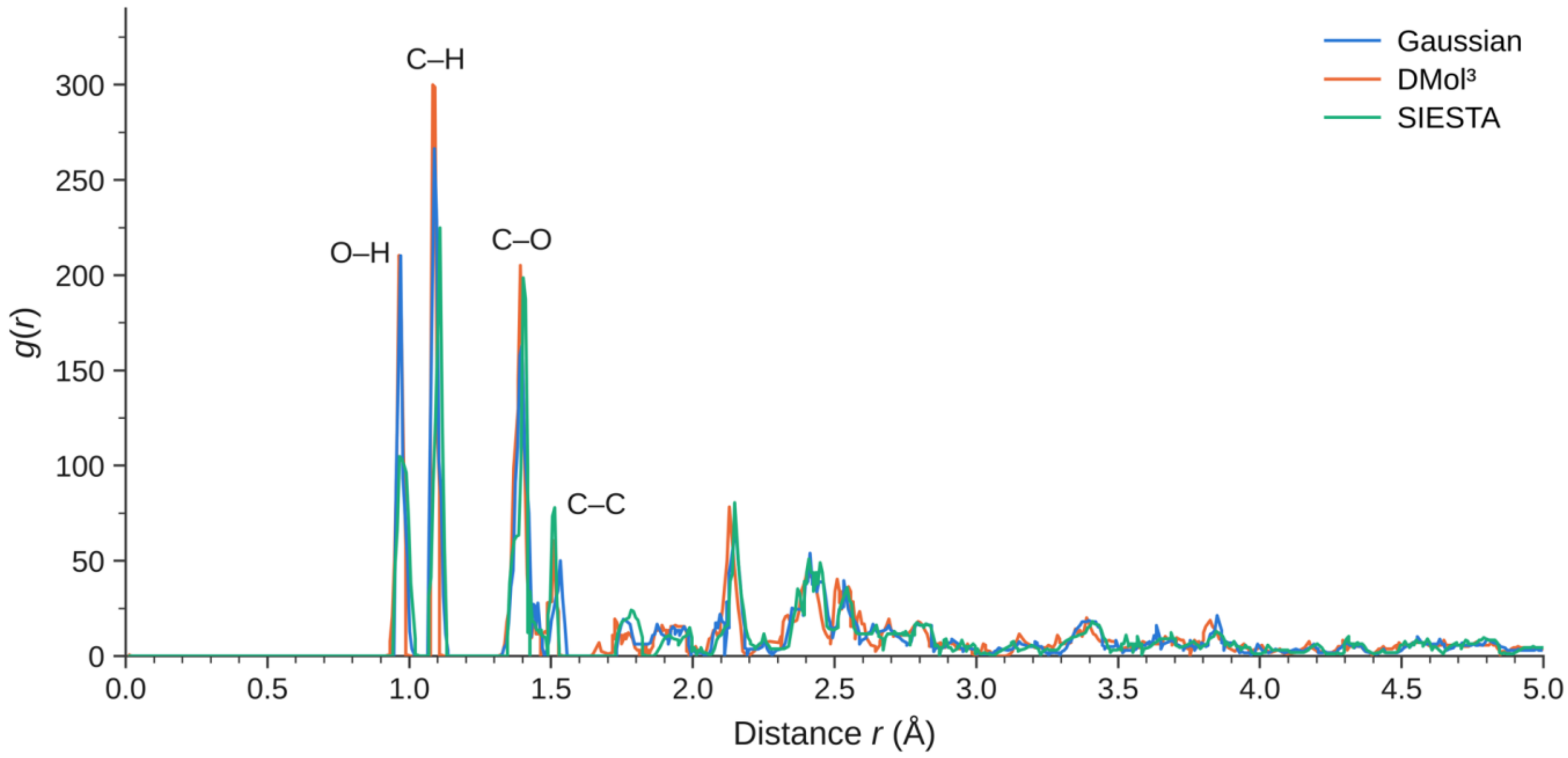

**Figure 1.** Results of the $g(r)$ calculations for the RF base unit from the Gaussian, $DMol^3$ and SIESTA software.

We then checked that the numerical representation of the RF base unit is in agreement with experiment by comparing an experimental infrared spectrum with the spectrum calculated from the harmonic vibrational frequencies (Section 2.2). The calculated and experimental spectra are compared in Figure 4. The low-frequency peaks show a rather organized system at the local scale, and all the main vibrational modes known in the RF structure, namely scissoring and stretching motions, have been identified. The superposition of the two spectra shows that the calculated molecular system is in agreement with the real system, which validates the numerical representation of the RF base unit obtained from the *ab initio* calculations.

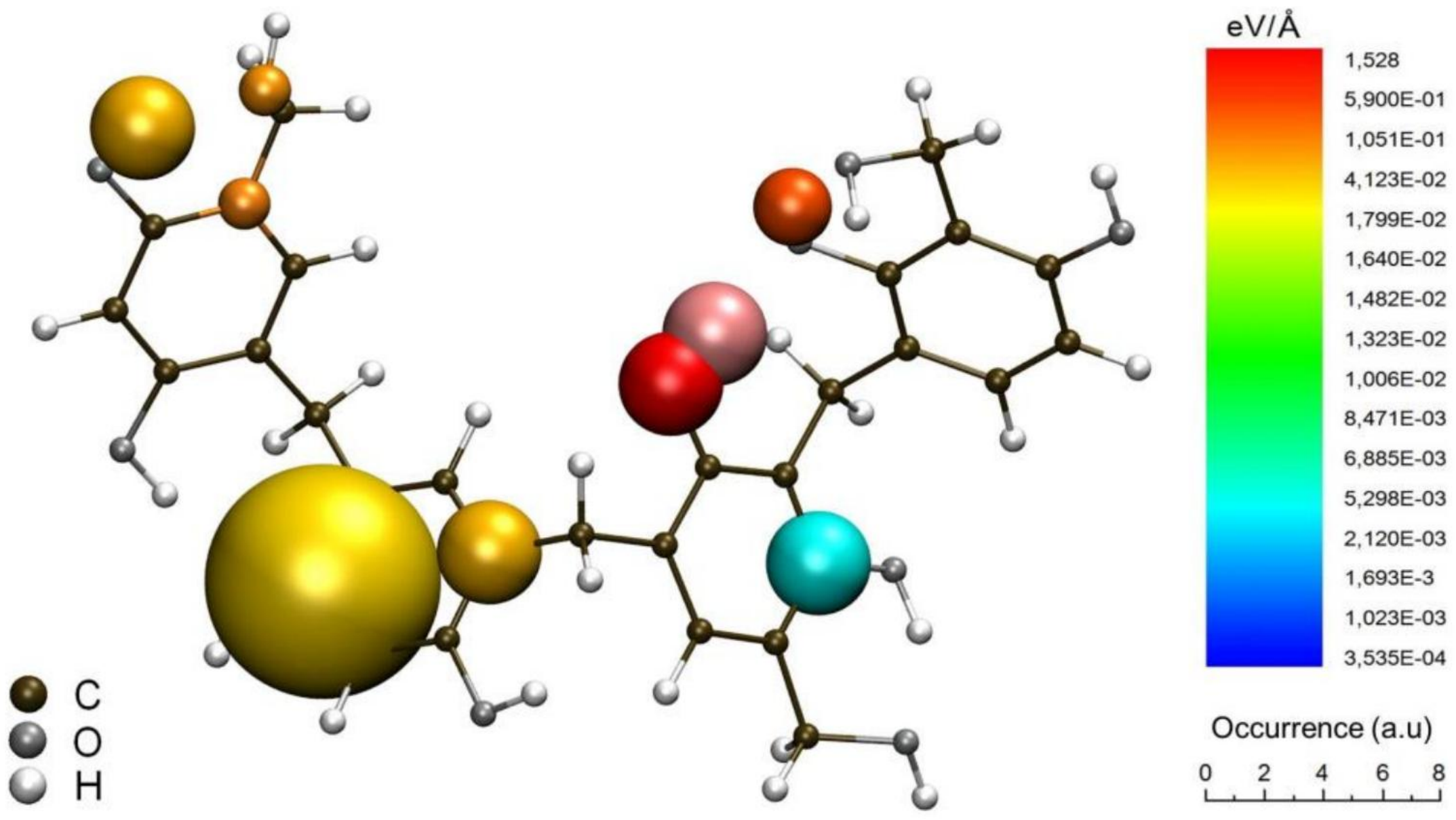

**Figure 2.** Largest interatomic forces during the SIESTA geometry optimization of the RF base unit. The colour of each sphere gives the magnitude of the force and its diameter the number of optimization steps in which that atom carried the largest force.

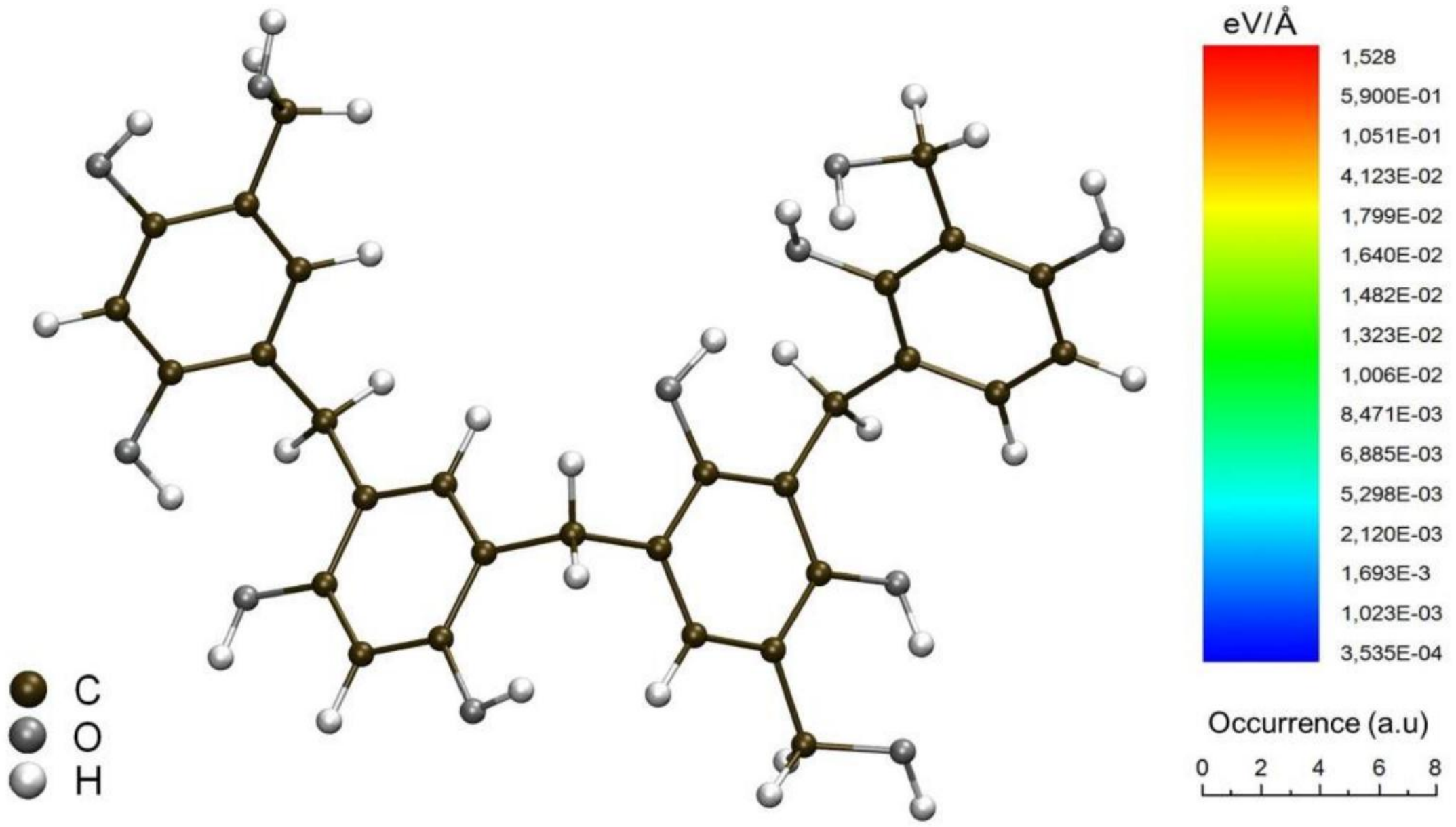

**Figure 3.** Forces on the atoms of the RF base unit at the end of the SIESTA geometry optimization. No sphere appears because the residual force on every atom is below the lowest value of the colour scale ($3.5 \times 10^{-4}$ eV Å$^{-1}$).

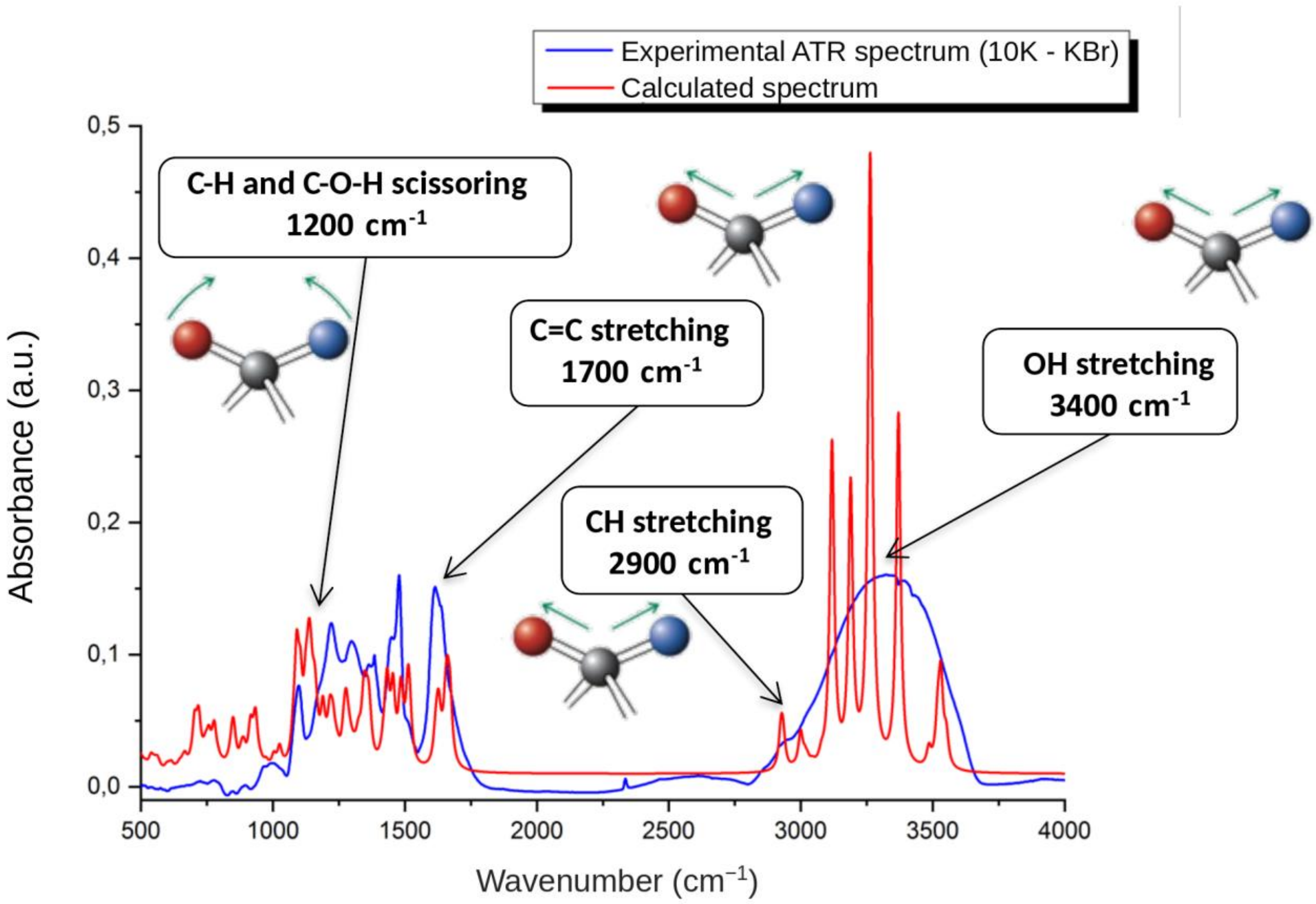

**Figure 4.** Theoretical and experimental infrared spectra of the RF base unit with the representation of the different vibrational modes.

### 3.2 Branched RF Units and Choice of the Force Field

The size of the system was then extended to several RF base units, while keeping the number of atoms small enough to remain within reasonable calculation times. The base unit, denoted N ($C_{30}H_{30}O_{11}$, 71 atoms), has three branching sites, because resorcinol itself has three reactive sites. Larger systems were built by linking additional base units to these sites through methylene ($CH_2$) bridges, as in a Novolac-type resin, in agreement with the synthesis of RF aerogels.

Each system is named after the fraction of the three branching sites of the central unit that are occupied:

– N+0.33: one site occupied, i.e. 2 base units ($C_{60}H_{58}O_{22}$, 140 atoms);
– N+0.66: two sites occupied, i.e. 3 base units ($C_{90}H_{86}O_{33}$, 209 atoms);
– N+1: all three sites occupied, i.e. 4 base units ($C_{120}H_{114}O_{44}$, 278 atoms).

For each size, several configurations were built, numbered (1) to (4), which differ in [the branching sites used / the spatial arrangement of the units]. The eight configurations studied are shown in Figure 5. These eight configurations were used to validate the force field for the MD calculations. To simulate the aggregation of clusters of several RF base units, the system must be much larger than a few hundred particles, which is not feasible in DFT. The four force fields of Section 2.3 were therefore used in geometry optimizations of the eight configurations with Forcite, and the results were compared with DFT geometry optimizations of the same configurations with Gaussian, the objective being to find the force field whose results are closest to those of DFT (Figure 6; Figure 7).

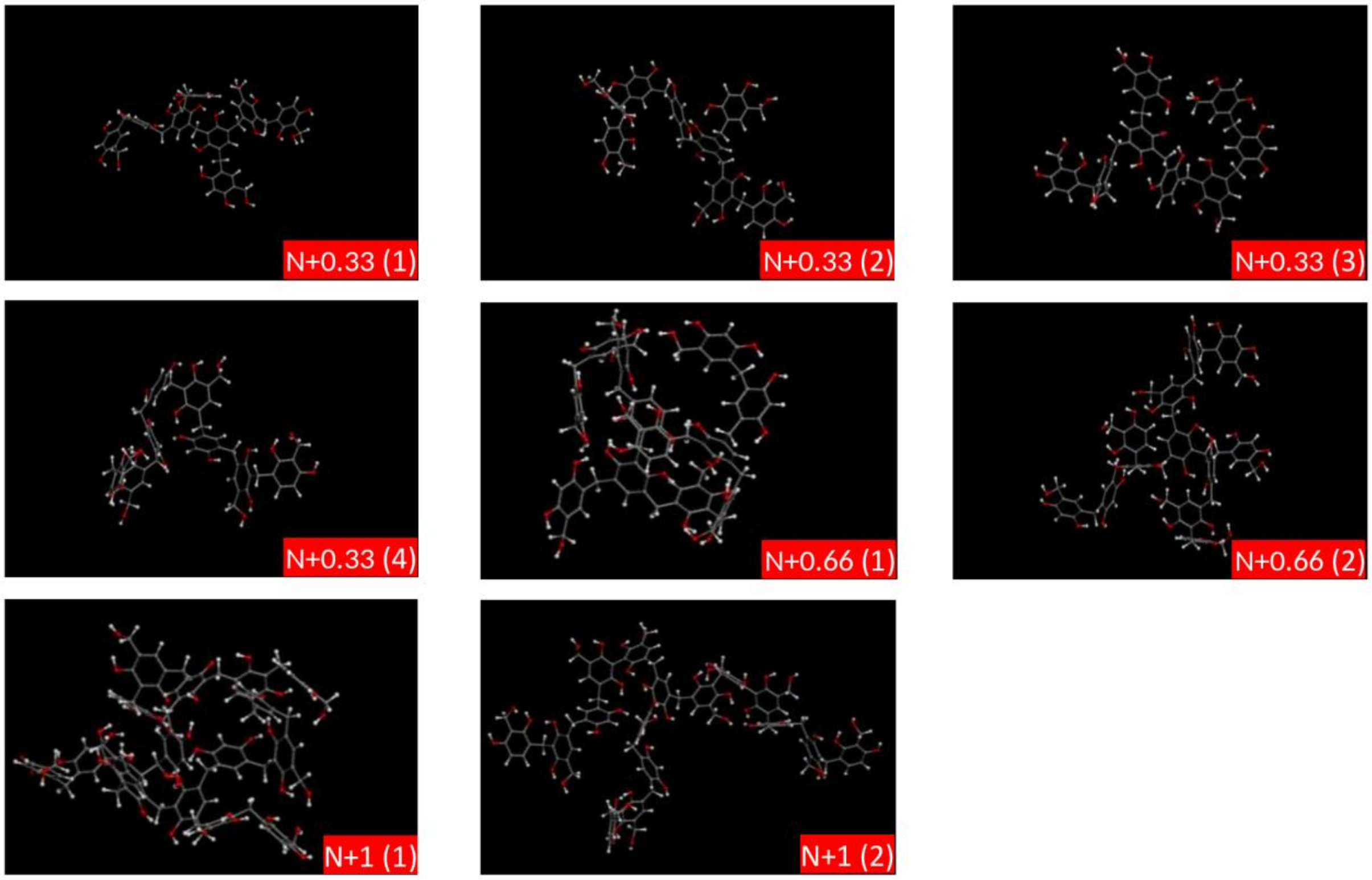

**Figure 5.** The eight configurations built from two (N+0.33), three (N+0.66) and four (N+1) RF base units; numbers in parentheses distinguish configurations of the same size. C, O and H atoms are shown in grey, red and white.

According to the DFT results from Gaussian, the larger the number of atoms in the system, the lower the energy per particle and therefore the more stable the system (Figure 6). The only force field to give this same trend is COMPASS II, which takes *ab initio* parameters into account. Moreover, for systems of the same size but of different conformations, the energy trend is also respected between the Gaussian and COMPASS II results (Figure 7). COMPASS II is therefore the force field best suited to our system and was retained for the rest of the calculations.

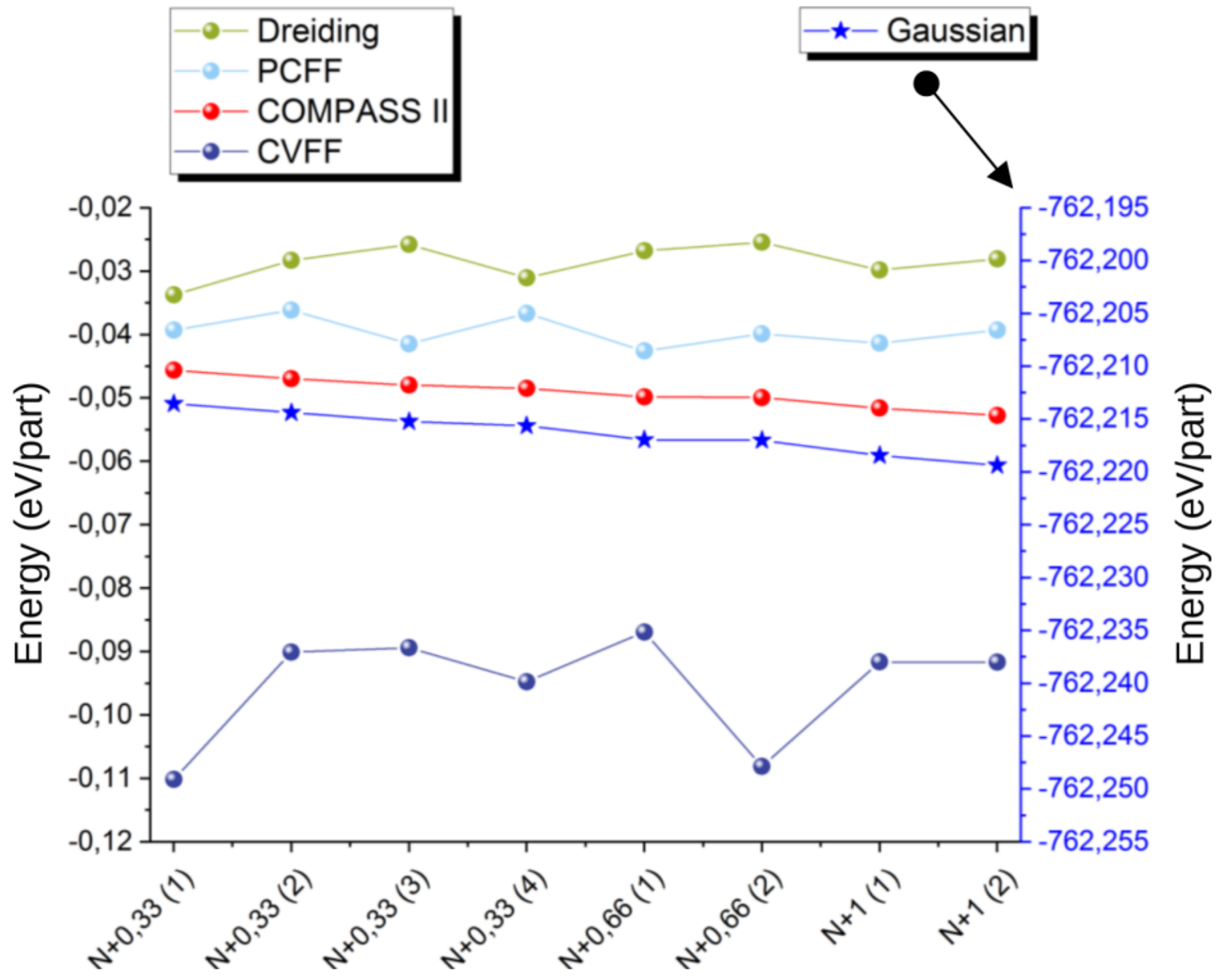

**Figure 6.** Comparison of the geometry optimization results between various classical-mechanics force fields (left-hand y-axis) and the results obtained with Gaussian in DFT (right-hand y-axis).

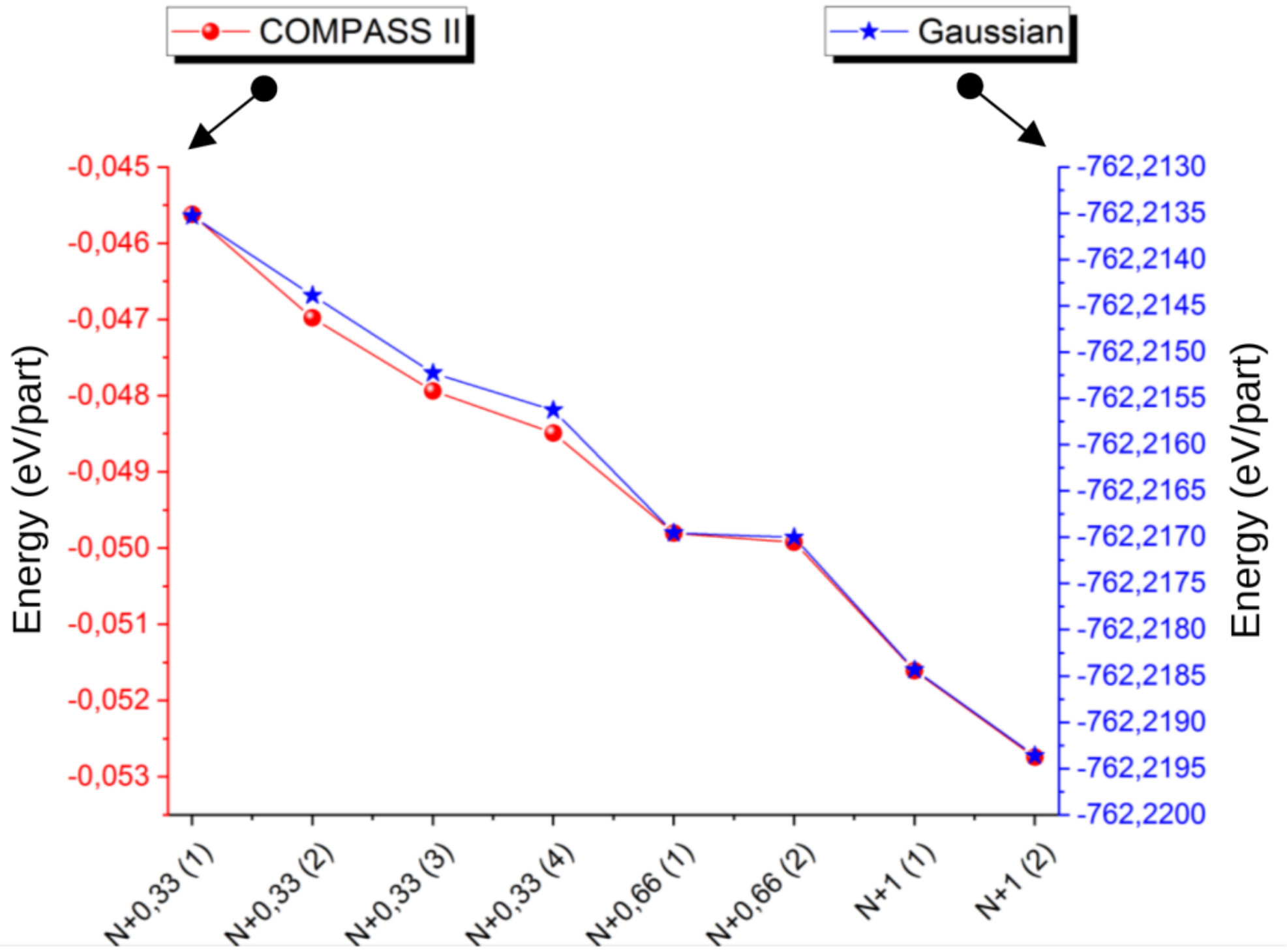

**Figure 7.** Comparison of the geometry optimization results with the classical-mechanics COMPASS II force field (left-hand y-axis) and the results obtained with Gaussian in DFT (right-hand y-axis).

### 3.3 Integration Time Step from the Symplecticity Test

The trajectory of several $N$+1 units was then studied, starting from the lowest-energy conformation, $N$+1 (2), an assembly of 4 RF base units linked via methylene bridges (278 atoms). To arrive at an aggregation phenomenon and build a system large enough to represent the real structure of the RF aerogel, we arbitrarily chose to study about 15,000 particles, which corresponds to 54 conformations of $N$+1 (2), i.e. 15,012 atoms in total. The time step was first chosen by checking the symplecticity of the method: the total energy of the system was followed in the microcanonical ensemble (*NVE*) over 0.1 ps, for time steps $\Delta t$ varying from 0.1 fs to 1 fs, the method being symplectic if the curves obtained for the different time steps fluctuate around the same mean value. This is the case for time steps up to 0.4 fs, whereas the curves for 0.5 fs and above deviate from this mean value (SI Section S2, Figures S1 and S2). The performance of the method was then checked via $\langle \Delta E_{\mathrm{tot}}^2 \rangle^{1/2}$, the root mean square of the fluctuations of the total energy, as a function of the integration step [43]: with the Velocity Verlet algorithm, this quantity must vary as $\Delta t^2$, which is verified for integration steps below 0.5 fs but not for larger ones (SI Figure S3). Since the largest time step that conserves the total energy minimizes the calculation time, the time step $\Delta t$ = 0.4 fs is the best suited to our system and was used for all the simulations.

### 3.4 Aggregation of the RF Units at Constant Volume

In the production phase, the 54 $N$+1 (2) units were placed randomly, with the Amorphous Cell module, in a cubic box with a side of 68 Å and periodic boundary conditions. The size of the box was chosen so that the density agrees with the experimental density of the RF skeleton found in the literature, 1.3 g cm$^{-3}$ [44]. An *NVT* relaxation of the 15,012 atoms was then performed with the Forcite module at room temperature (300 K, Nosé–Hoover thermostat), with a time step of 0.4 fs and a total simulation time of 1.2 ns, long enough for the system to reach an equilibrium state (Figure 8). The different base units clearly tend to aggregate: at t = 0 ps, the RF molecules are dispersed throughout the volume of the simulation box; at t = 10 ps, a few clusters begin to appear; at t = 50 ps, a single cluster forms towards the centre of the box, which continues until about t = 300 ps; up to t = 1200 ps, the system hardly evolves anymore and the aggregation process is complete. The potential energy, the temperature and the total energy of the system during this simulation are shown in Figure 9.

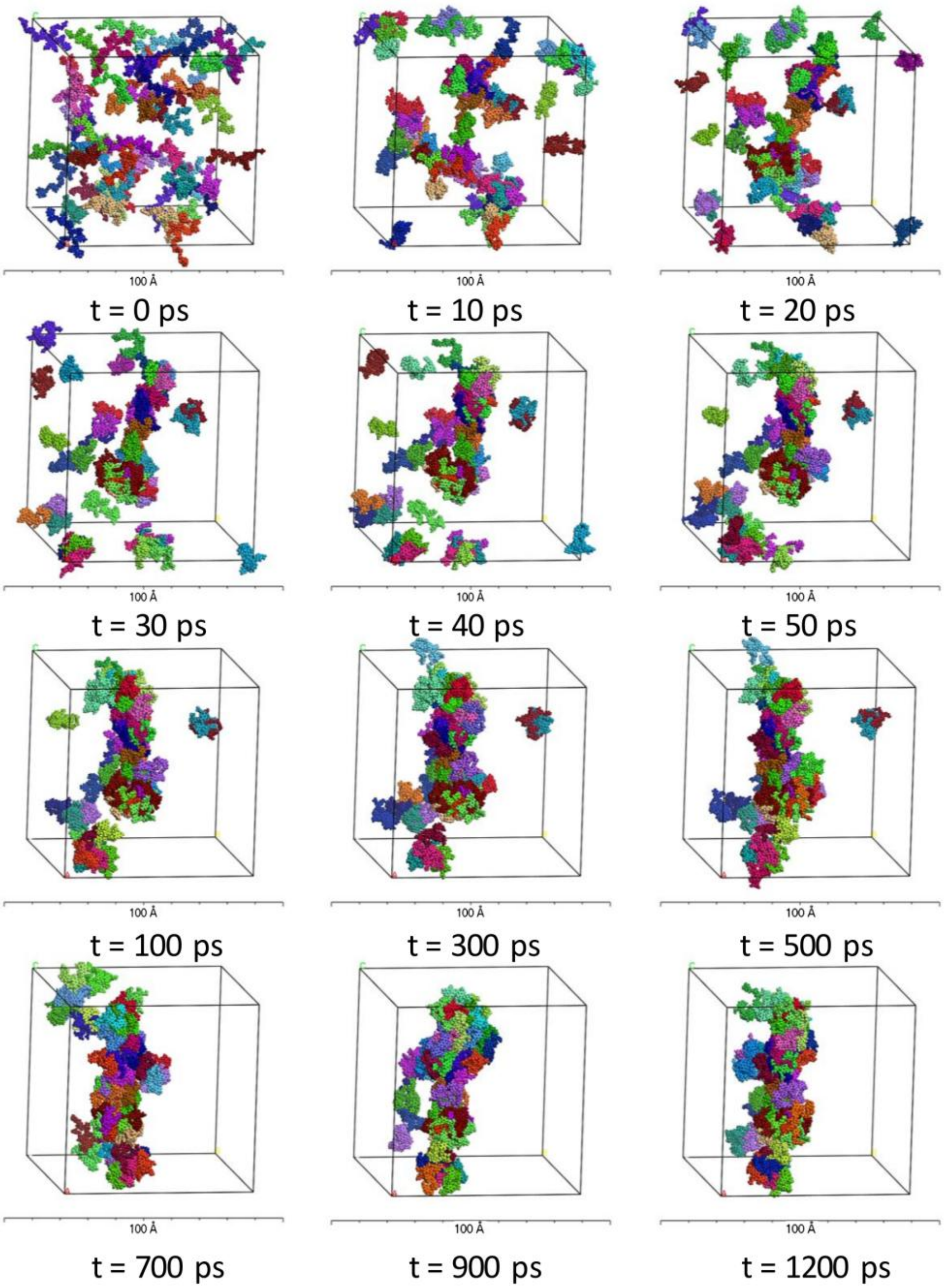

**Figure 8.** Evolution of the structure of the system during an *NVT* simulation of 1200 ps.

The potential energy decreases very rapidly up to about 100 ps, then more slowly up to about 300 ps, and finally stabilizes up to 1.2 ns. The temperature remains stable at 300 K, which demonstrates that the Nosé–Hoover thermostat is perfectly suited to this system, and the total energy, the sum of the kinetic energy (expressed here as temperature) and of the potential energy, follows the same behavior as the potential energy. These energy curves agree with the trajectory of Figure 8, and a relaxed system is obtained at the end of the 1.2 ns of *NVT* simulation.

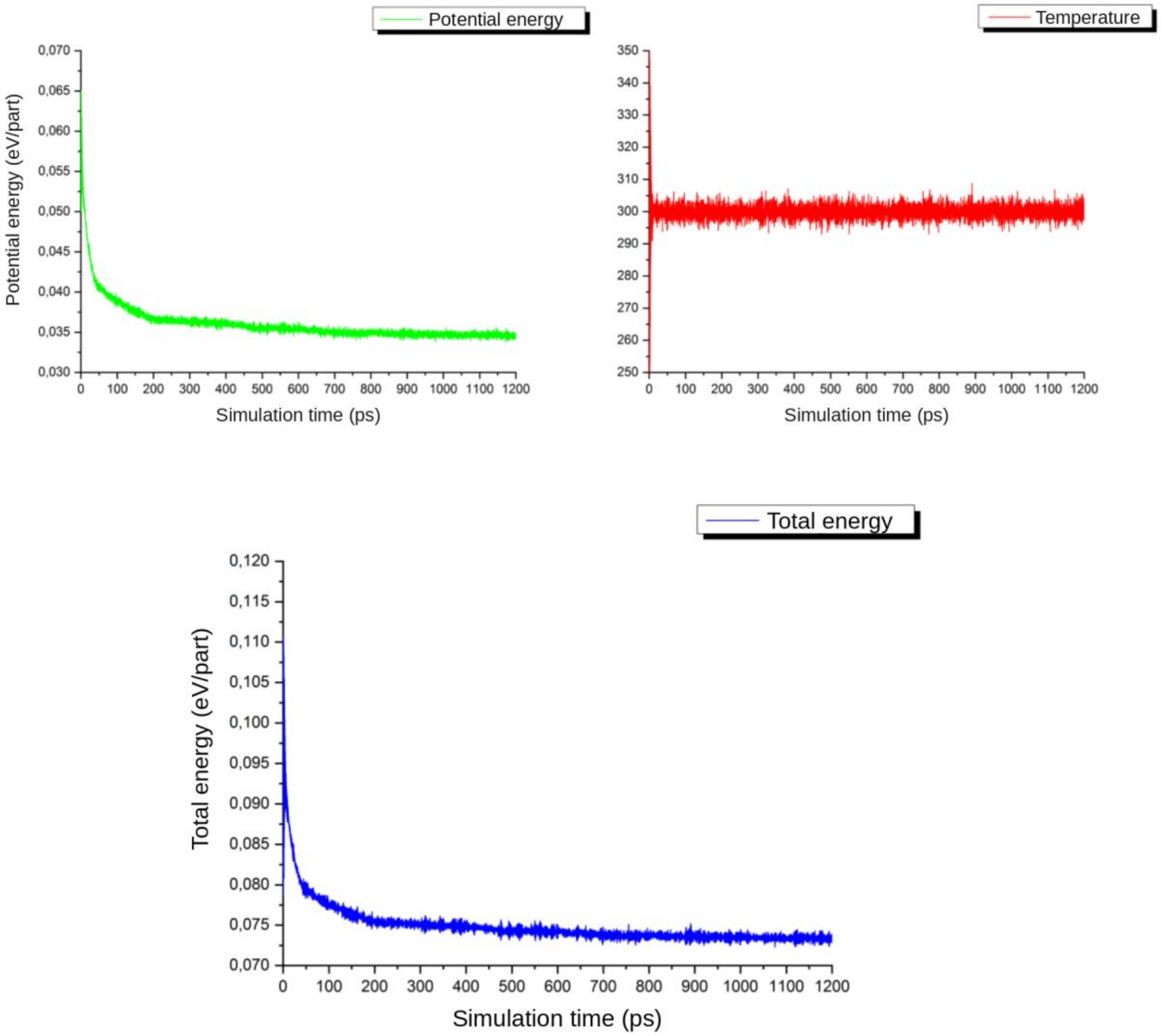

**Figure 9.** Evolution of the potential energy, the temperature and the total energy of the system during the *NVT* simulation.

### 3.5 Equilibrium Density and Comparison with United-Atom Models

The *NVT* simulation yields an RF aerogel structure in the form of a single cluster, but since the volume of the box is fixed in the canonical ensemble, the relaxed system is certainly not at its equilibrium density. An additional step was therefore carried out in the isobaric–isothermal ensemble (*NPT*), with the pressure and the temperature fixed by the Berendsen barostat and the Nosé–Hoover thermostat, respectively, so that the volume and the energy of the system can evolve. Starting from the last structure of the *NVT* simulation (after 1.2 ns), the density, the potential energy, the temperature and the total energy were followed over 550 ps with unchanged calculation parameters (Figure 10). The system densifies and reaches an equilibrium state after only a few picoseconds. Despite the use of the same calculation parameters, the total energy is much lower (about −0.035 eV/part in *NPT* versus about +0.075 eV/part in *NVT*), which demonstrates that the densified system is much more stable. The Nosé–Hoover thermostat maintains the system at the equilibrium temperature of 300 K, and the equilibrium density is reached for a value of 1.296 g cm$^{-3}$, in agreement with the density of the RF skeleton found in the literature, namely 1.3 g cm$^{-3}$ [44]. The cluster now occupies the whole volume of the periodic box (Figure 11).

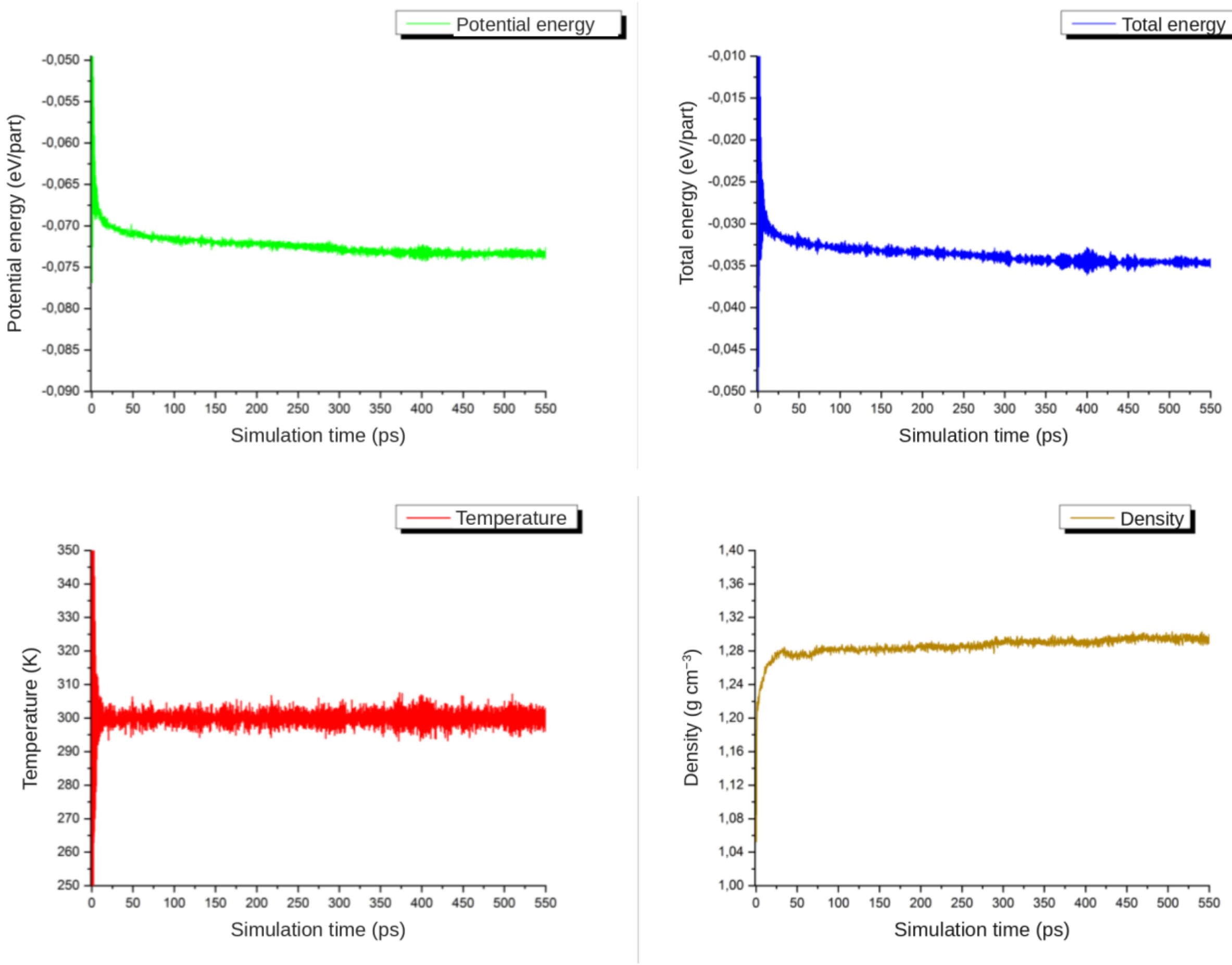

**Figure 10.** Evolution of the potential energy, the temperature, the total energy and the density of the system during the *NPT* simulation.

The modelled system therefore reproduces the experimental skeletal density, and its base unit reproduces the experimental infrared spectrum. These validations concern selected observables; a comparison with X-ray or neutron total scattering data would test the structure of the skeleton more directly. The model can now serve as a starting point for the study of doping processes or heat treatments, with a view to improving the thermoelectric properties. These results can be compared with the united-atom MD models of resorcinol resins recently reported by Izumi et al. 2024 [12]. In their approach, a network of 3,000 resorcinol monomers is built by successive pseudo-reactions whose probability depends on the interatomic distance and on the relative orientation of the resorcinol rings; for the resin obtained by addition–condensation of resorcinol with formaldehyde, the density reaches 1.23 g $cm^{-3}$, and the calculated elastic modulus and Poisson's ratio (4.0 GPa and 0.35) agree with those of typical phenolic resins. The two approaches are complementary. The united-atom description, without explicit hydrogen atoms or electrostatic interactions, makes it possible to follow the cross-linking reactions and the gel point in large systems, whereas the all-atom COMPASS II model used here, includes explicit hydrogen atoms and electrostatic interactions, and hence hydrogen bonding, is validated against DFT and the experimental infrared spectrum, and reaches a density (1.296 g $cm^{-3}$) close to the experimental skeletal density of RF aerogels. Conversely, no covalent bonds are formed between the 54 $N$+1 (2) units during our simulations: these units, each branched internally through methylene bridges, assemble through non-bonded interactions only. The equilibrated structure therefore represents a physically aggregated RF skeleton rather than a fully cross-linked network, and cross-linking the aggregated units, for example with a geometry-based algorithm such as that of Izumi et al. 2024 [12], would be a natural extension of this work.

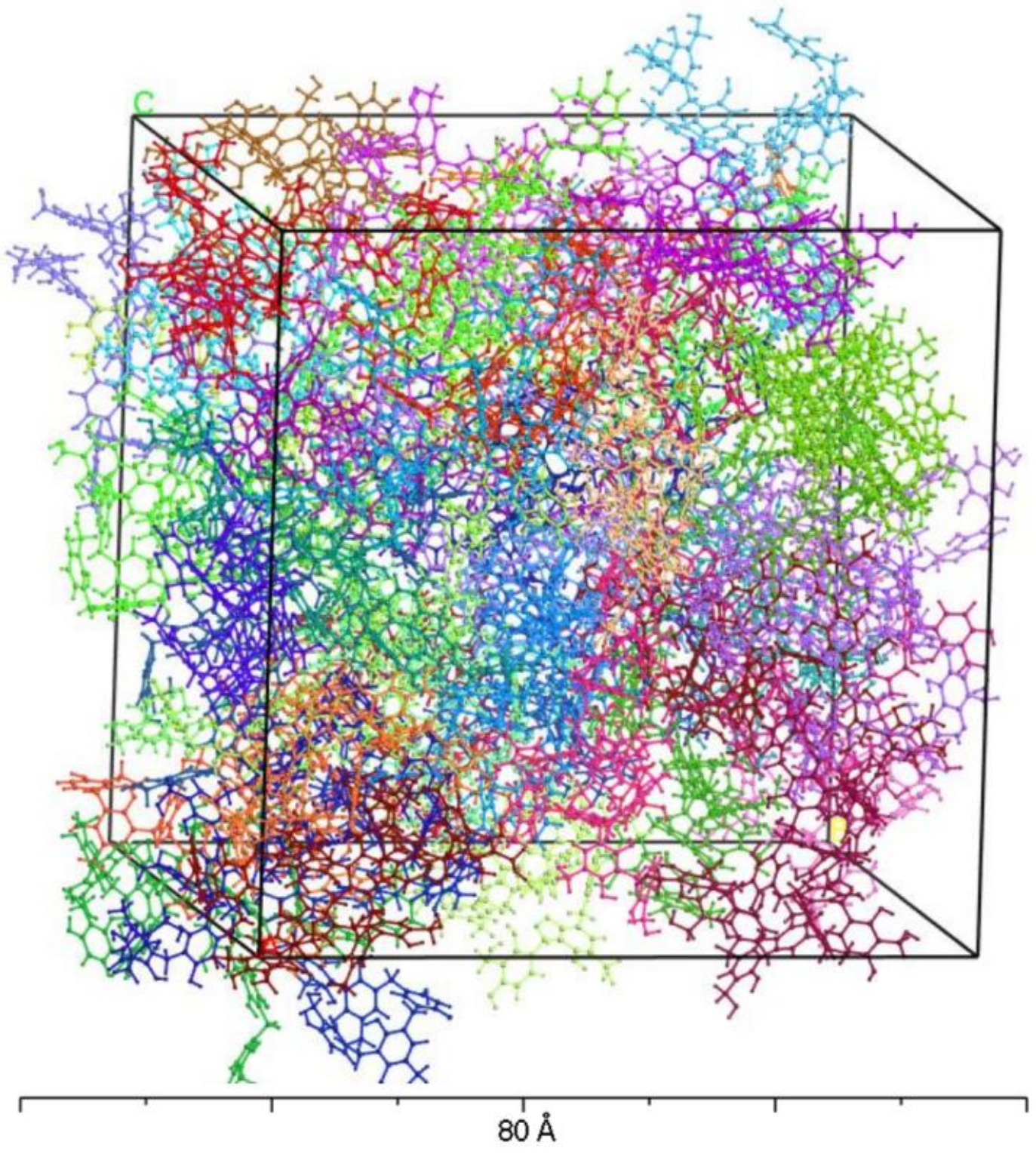

**Figure 11.** Densified system in its equilibrium state, at the end of the *NPT* simulation.

### 3.6 Incorporation of Iron into the RF Skeleton

To improve the thermoelectric properties, doping could induce conduction paths and thus improve the electronic properties of the material. In a first study, we chose iron particles, which are an excellent candidate for improving the electrical conductivity of the matrix in which they are located, owing to their excellent electrical conductivity, their ease of synthesis and their low cost [45–47]. The chosen doping level is 5% by mass, which is a reasonable level to begin the study of the influence of iron particles on the electrical properties of the RF material; other doping levels could be compared in future simulations. The 109 iron atoms required (Table 2) were inserted randomly, by the same Monte Carlo procedure with the Amorphous Cell module, into the relaxed structure obtained at the end of the *NPT* simulation of Section 3.5. Because this addition modifies the equilibrium density of the system, a new *NPT* simulation of 1 ns was carried out with the same parameters. The iron atoms, inserted as isolated atoms, were described with the COMPASS II force field, like the RF skeleton, as neutral atoms (charge 0 e) interacting with the skeleton and with each other through non-bonded terms only, with 9-6 Lennard-Jones parameters $\varepsilon = 0.013$ kcal mol$^{-1}$ and $r_0 = 2.91$ Å.

**Table 2.** Main parameters used for the 5 wt% Fe-doped RF model.

| Quantity | Value |
|---|---|
| Molar mass of one *N*+1 (2) unit ($C_{120}H_{114}O_{44}$) | $M_{N+1(2)}$ = 2260.188 g mol$^{-1}$ |
| Molar mass of the undoped system (54 *N*+1 (2) units) | $M_1$ = 122,050.15 g mol$^{-1}$ |
| Doping of 5% by mass | $N_{Fe} = (0.05 \times M_1)/M_{Fe}$ = 109 |
| Molar mass of the doped system | $M_{total} = M_1 + (109 \times M_{Fe})$ = 128,137.47 g mol$^{-1}$ |
| Calculated theoretical density of the undoped system | $d_{theo}$ = 1.296 g cm$^{-3}$ |
| Expected theoretical density of the doped system | $d = (M_{total}/M_1) \times d_{theo}$ = 1.361 g cm$^{-3}$ |

During the 1 ns *NPT* simulation, the doped system reaches an equilibrium state after only a few tens of picoseconds, and the Nosé–Hoover thermostat still maintains it at 300 K (Figure 12). The equilibrium density, of the order of 1.36 g cm$^{-3}$, is in agreement with the value expected for this doped system (Table 2), so that the system can be considered at equilibrium after 1 ns of simulation. As in the undoped system (Figure 11), the RF cluster occupies the entire volume of the periodic box, and the iron particles (yellow spheres) tend to penetrate the amorphous RF network, which is, a priori, a good thing for doping purposes (Figure 13). Since the density increases in proportion to the added mass (Table 2), the volume of the box is essentially unchanged: the iron atoms occupy free volume within the RF network rather than expanding it.

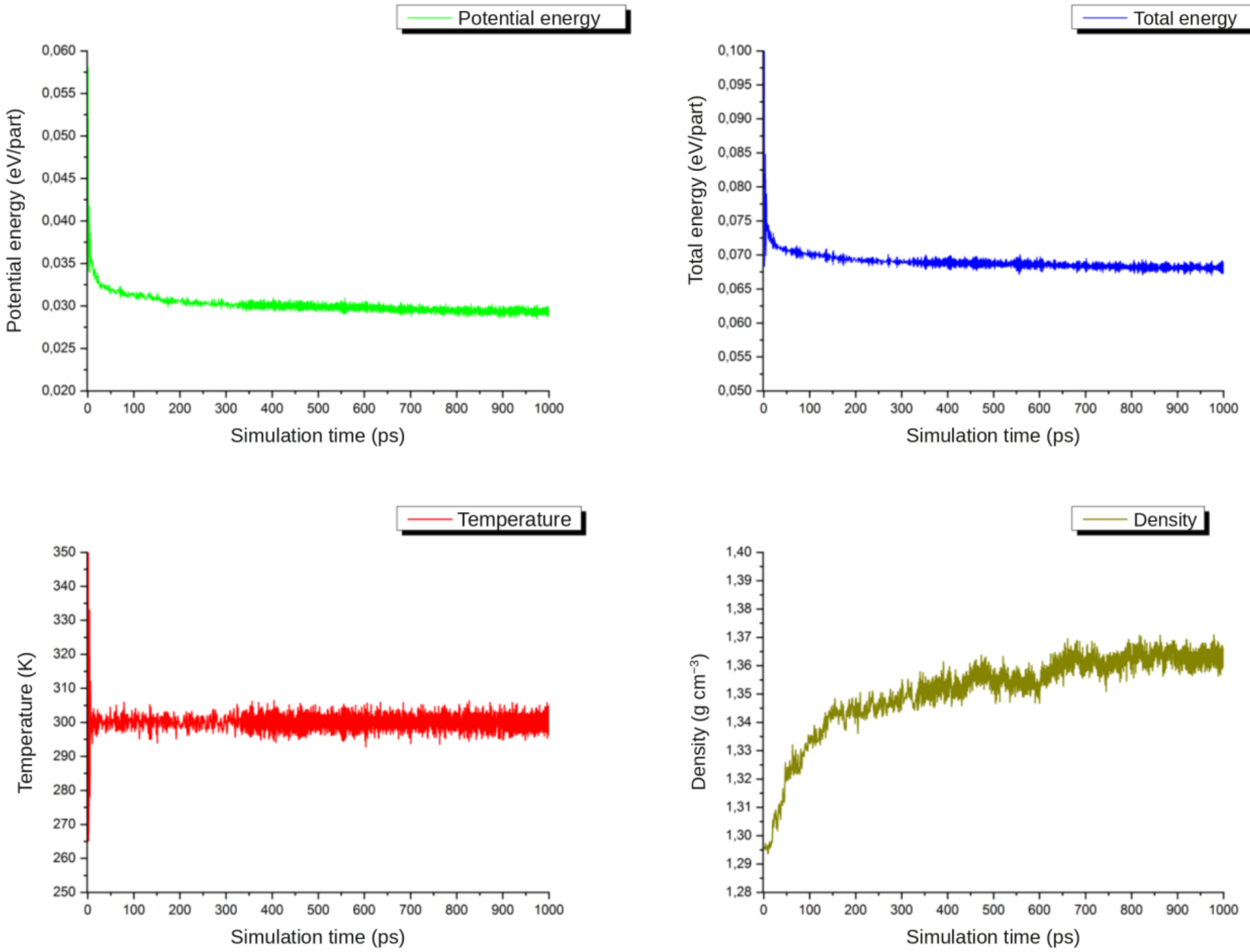

**Figure 12.** Evolution of the potential energy, the temperature, the total energy and the density of the iron-doped system during the *NPT* simulation.

To determine the arrangement of the iron atoms in the RF gel, we calculated, from the last configuration of the doped system, the partial radial distribution function between the iron atoms and the carbon atoms of the gel, $RDF(r) = 4\pi r^2 \rho_0 c_C g_{FeC}(r)$ (SI Eq. S4). Unlike the pair distribution function g(r) (Section 3.1), which tends to 1 at large distances, the RDF gives the number of atoms per unit distance around a given atom, so that its integral gives coordination numbers (SI Section S1.2, Eqs. S1–S3). The resulting Fe–C RDF is shown in Figure 14. A well-defined peak appears at about 3 Å, together with a few less well-defined peaks between 4 Å and 6 Å. The peak at 3 Å can be attributed to the mean distance between the iron atoms within the RF structure and the carbon atoms, whereas the peaks at 4 Å and 6 Å could correspond to the iron–carbon distance for iron atoms still located at the surface of the RF gel.

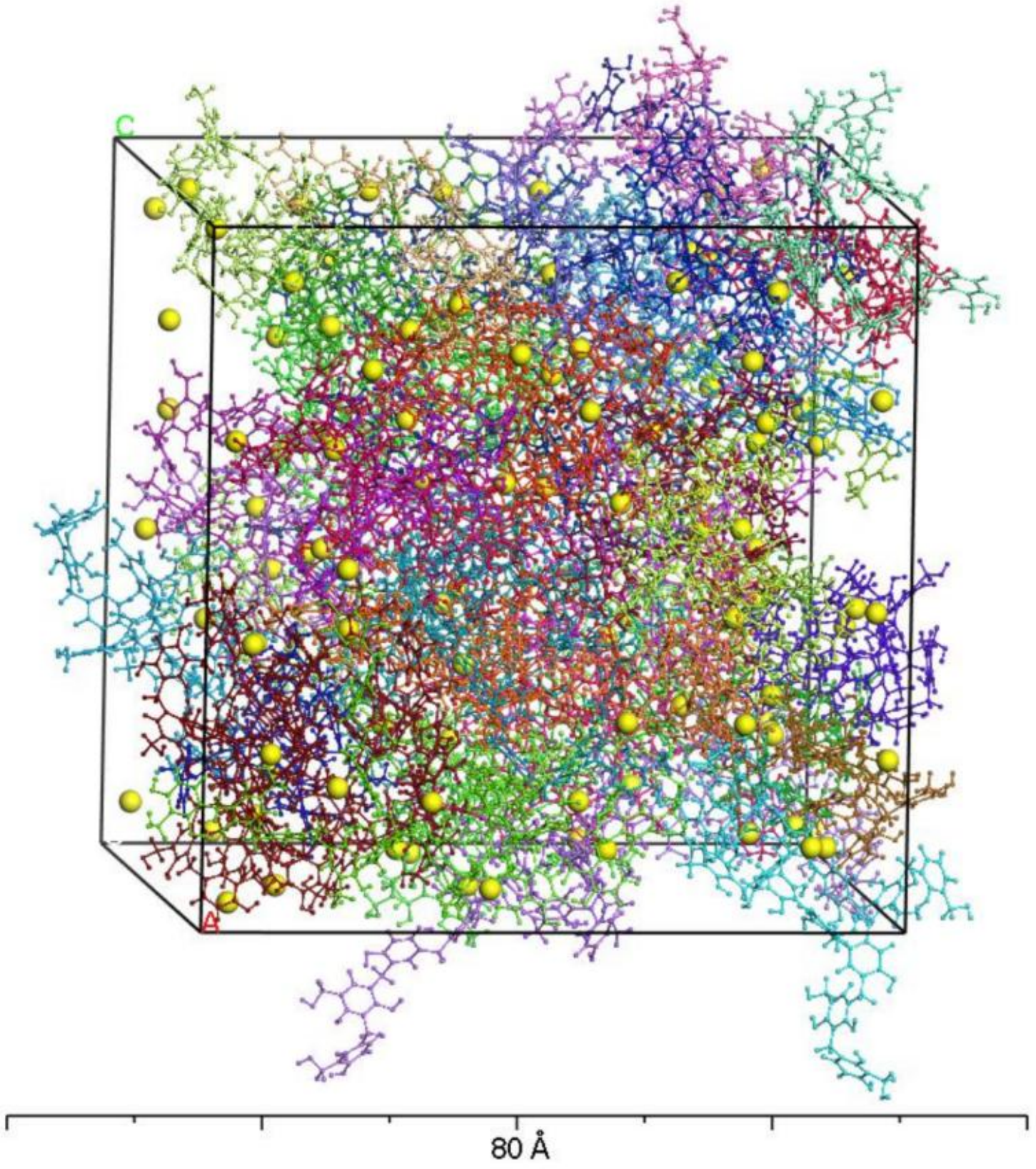

**Figure 13.** Iron-doped system densified in its equilibrium state, at the end of the *NPT* simulation.

We then followed the displacement of each iron particle during the *NPT* simulation, together with the number of its interactions with its nearest neighbors within a radius of 5 Å. In view of Figure 14, this is the cut-off radius for which the arrangement of an iron particle in the network of the RF gel is certain, in agreement with the literature [48, 49]; moreover, the integral of the RDF (SI Eq. S3) shows that an iron atom present in the network of the gel interacts with, on average, 20 carbon atoms within this radius. This value is close to the number of carbon atoms expected within 5 Å of an atom fully surrounded by the skeleton at its mean carbon density (about 22, for 6,480 carbon atoms in about 1.56 × $10^5$ Å$^3$), whereas an atom at the surface would have roughly half as many. Every 50 ps, the coordinates $(x_C, y_C, z_C)$ of the carbon atoms and $(x_{Fe}, y_{Fe}, z_{Fe})$ of the iron atoms were recorded, and each iron atom interacting with at least 20 carbon atoms within a radius of 5 Å was considered to be at its equilibrium position, within the structure of the RF gel (Figure 15). Although the system reached an equilibrium state after 1 ns of *NPT* simulation, not all the iron particles entered the RF network; some are perhaps only at the surface of the structure. The fraction of embedded iron particles increases from 63.3% (69 of 109) immediately after insertion to 70.6% (77) after 800 ps, and then remains stable up to 1 ns (Figure 15). These results are encouraging with regard to a doping process, but complementary studies must be carried out, for example varying the initial doping level in order to guarantee a homogeneous distribution of the fillers throughout the RF matrix and thus reach a percolation threshold.

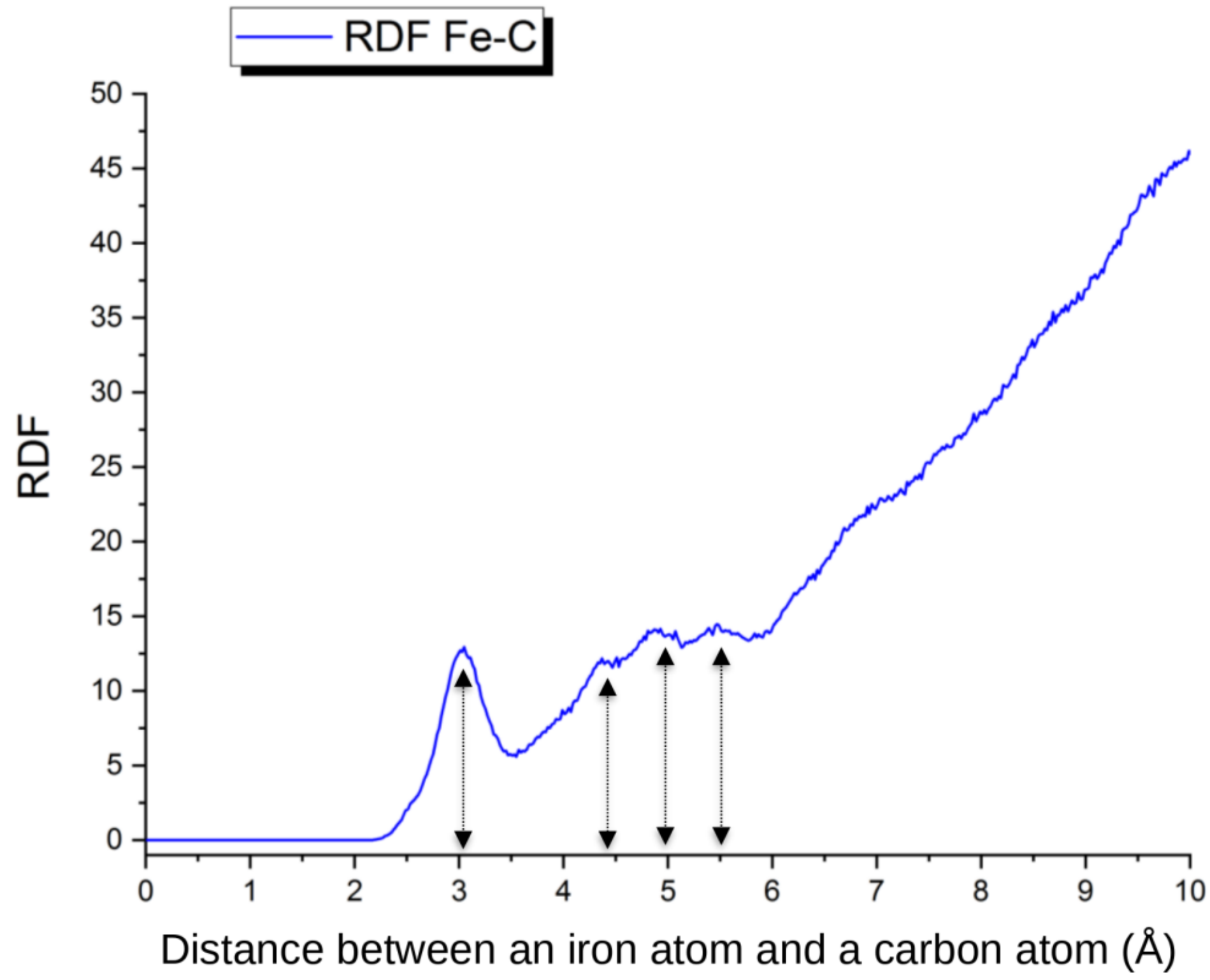

**Figure 14.** Fe-C radial distribution function within a cut-off radius of 10 Å at the end of the 1 ns *NPT* simulation.

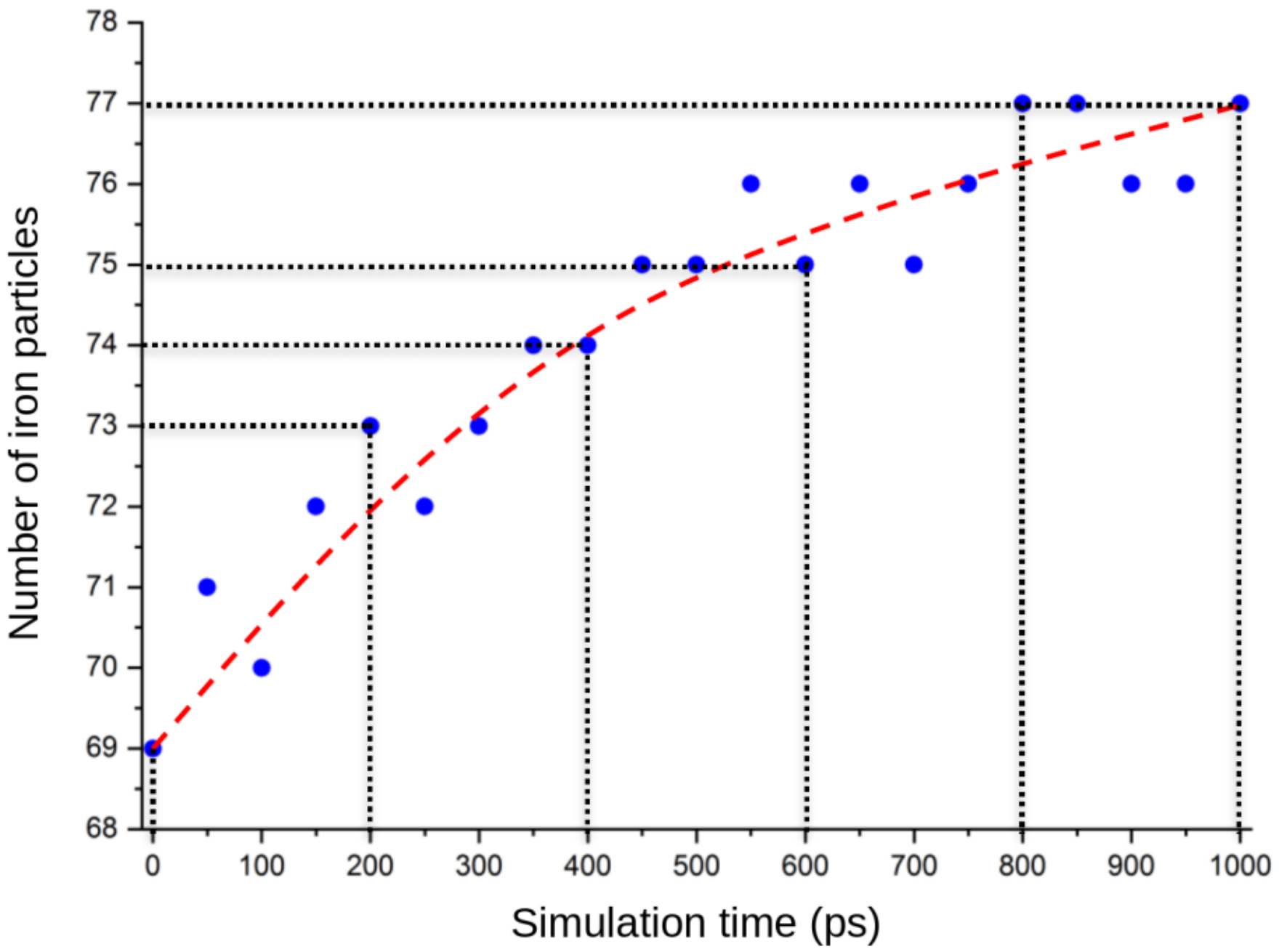

**Figure 15.** Number of iron particles having at least 20 interactions with a carbon atom within a radius of 5 Å during the *NPT* simulation.

## 4. Conclusions

We built an aggregated all-atom numerical representation of the RF aerogel from scratch, starting from the resorcinol ($C_6H_4(OH)_2$) and formaldehyde ($CH_2O$) precursors, which complements the recent united-atom models of resorcinol resins [12]. The RF base unit was first optimized by DFT with Gaussian and DMol$^3$ (B3LYP functional) and SIESTA (GGA-PBE functional), with free boundary conditions; the three codes gave similar structures, as confirmed by their pair distribution functions, and the calculated infrared spectrum agreed with the experimental one, which validates the numerical representation of the RF base unit. Structures of several base units, N+0.33 (140 atoms), N+0.66 (209 atoms) and N+1 (278 atoms), were then optimized both by DFT and with several classical force fields, which led to the selection of the COMPASS II force field of the Forcite module of Materials Studio. The system was then enlarged to 54 N+1 units, i.e. 15,012 particles, placed randomly in a cubic periodic box of 68 Å side, consistent with the experimental density. A symplecticity test in the microcanonical ensemble (*NVE*) defined the time step best suited to the integration of Newton's equations (0.4 fs). In the canonical ensemble (*NVT*), the RF units aggregated after only a few tens of picoseconds, and in the isobaric–isothermal ensemble (*NPT*) the system converged towards an equilibrium density of 1.296 g cm$^{-3}$, in agreement with the experimental density (1.3 g cm$^{-3}$). In this model, the units assemble through non-bonded interactions only; cross-linking them covalently, as recently done for resorcinol resins with a united-atom description [12], is a natural next step. Finally, a first doping study was carried out by introducing 5 wt% of iron particles (109 atoms) into the equilibrated system and simulating it for 1 ns in the *NPT* ensemble. The radial distribution function revealed a peak at 3 Å, corresponding to the iron–carbon distance for iron particles within the network of the RF gel, and the fraction of iron particles inside the amorphous RF network increased from 63% after insertion to 71% after 1 ns, a promising result in the context of doping. The quality of the dispersion of the fillers nevertheless remains difficult to evaluate, and complementary studies, including different doping levels, must be carried out. Two extensions follow from this work. First, the thermoelectric properties of the doped skeleton can now be calculated from configurations extracted from the MD trajectories, with transport methods suited to disordered materials. Other fillers, such as carbon nanotubes, reduced graphene oxide or metal salts, can be compared in the same way. Second, beyond thermoelectrics, once its units are cross-linked, the model could be extended to predict the skeleton's thermal conductivity and elastic constants, which are key inputs for modelling RF aerogels as thermal insulators. It could also serve as the starting structure for simulating the pyrolysis of RF gels into carbon and metal-doped carbon aerogels for energy storage and catalysis [3, 45, 47].

## Acknowledgments

Shoeb Athar and Jeremy Guazzagaloppa acknowledge the financial support from the Hutchinson SA (Châlette-sur-Loing, France).

## References

[1] R.W. Pekala, Organic aerogels from the polycondensation of resorcinol with formaldehyde, Journal of Materials Science, 24 (1989) 3221–3227.

[2] X. Lu, M.C. Arduini-Schuster, J. Kuhn, O. Nilsson, J. Fricke, R.W. Pekala, Thermal conductivity of monolithic organic aerogels, Science, 255 (1992) 971–972.

[3] S.A. Al-Muhtaseb, J.A. Ritter, Preparation and properties of resorcinol–formaldehyde organic and carbon gels, Advanced Materials, 15 (2003) 101–114.

[4] G.J. Snyder, E.S. Toberer, Complex thermoelectric materials, Nature Materials, 7 (2008) 105–114.

[5] H. Shirakawa, E.J. Louis, A.G. MacDiarmid, C.K. Chiang, A.J. Heeger, Synthesis of electrically conducting organic polymers: halogen derivatives of polyacetylene, (CH)x, Journal of the Chemical Society, Chemical Communications, (1977) 578–580.

[6] B. Russ, A. Glaudell, J.J. Urban, M.L. Chabinyc, R.A. Segalman, Organic thermoelectric materials for energy harvesting and temperature control, Nature Reviews Materials, 1 (2016) 16050.

[7] S. Athar, A. Bergonzoni, J. Guazzagaloppa, C. Huillet, P. Jund, Ab initio and experimental investigation of thermoelectric properties in a silica-based superinsulating material, Journal of Physical Chemistry C, 127 (2023) 9973–9980.

[8] S. Athar, J. Guazzagaloppa, F. Boyrie, C. Huillet, P. Jund, Carbogels for sustainable and scalable thermoelectric applications, npj Clean Energy, 1 (2025) 3.

[9] A. Izumi, T. Nakao, M. Shibayama, Atomistic molecular dynamics study of cross-linked phenolic resins, Soft Matter, 8 (2012) 5283–5292.

[10] J.D. Monk, J.B. Haskins, C.W. Bauschlicher Jr., J.W. Lawson, Molecular dynamics simulations of phenolic resin: construction of atomistic models, Polymer, 62 (2015) 39–49.

[11] M.A. Bone, T. Macquart, I. Hamerton, B.J. Howlin, A novel approach to atomistic molecular dynamics simulation of phenolic resins using symthons, Polymers, 12 (2020) 926.

[12] A. Izumi, Y. Shudo, K. Hagita, Y. Itoh, Geometry-based cross-linking algorithm for modeling resorcinol resins through united-atom molecular dynamics simulations, Polymer, 293 (2024) 126677.

[13] D. Systèmes, Materials Studio, https://www.3dsbiovia.com/products/collaborative-science/biovia-materials-studio/, (2017).

[14] W. Hehre, W. Lathan, R. Ditchfield, M. Newton, J. Pople, Gaussian 70 (Quantum Chemistry Program Exchange, in, Program, 1970.

[15] B. Delley, D.E. Ellis, A.J. Freeman, E. Baerends, D. Post, Binding energy and electronic structure of small copper particles, Physical Review B, 27 (1983) 2132.

[16] B. Delley, An all-electron numerical method for solving the local density functional for polyatomic molecules, The Journal of Chemical Physics, 92 (1990) 508–517.

[17] J.M. Soler, E. Artacho, J.D. Gale, A. García, J. Junquera, P. Ordejón, D. Sánchez-Portal, The SIESTA method for ab initio order-N materials simulation, Journal of Physics: Condensed Matter, 14 (2002) 2745.

[18] R. Robles, J. Izquierdo, A. Vega, L. Balbás, All-electron and pseudopotential study of the spin-polarization of the V (001) surface: LDA versus GGA, Physical Review B, 63 (2001) 172406.

[19] J. Izquierdo, A. Vega, L. Balbás, D. Sánchez-Portal, J. Junquera, E. Artacho, J.M. Soler, P. Ordejón, Systematic ab initio study of the electronic and magnetic properties of different pure and mixed iron systems, Physical Review B, 61 (2000) 13639.

[20] M. Frisch, G. Trucks, H.B. Schlegel, G. Scuseria, M. Robb, J. Cheeseman, G. Scalmani, V. Barone, B. Mennucci, G. Petersson, Gaussian 09, revision a. 02, gaussian, Inc., Wallingford, CT, 200 (2009) 28.

[21] E.B. Wilson, J.C. Decius, P.C. Cross, Molecular Vibrations: The Theory of Infrared and Raman Vibrational Spectra, McGraw-Hill, New York, 1955.

[22] P. Hermet, Introduction à la théorie de la fonctionnelle de la densité et application aux matériaux thermoélectriques, http://gdr-thermoelectricite.cnrs.fr/Contributions-Montpellier2017/Hermet-GIS2017.pdf, (2019).

[23] S.L. Mayo, B.D. Olafson, W.A. Goddard, DREIDING: a generic force field for molecular simulations, Journal of Physical Chemistry, 94 (1990) 8897–8909.

[24] P. Dauber-Osguthorpe, V.A. Roberts, D.J. Osguthorpe, J. Wolff, M. Genest, A.T. Hagler, Structure and energetics of ligand binding to proteins: Escherichia coli dihydrofolate reductase–trimethoprim, a drug–receptor system, Proteins: Structure, Function, and Bioinformatics, 4 (1988) 31–47.

[25] H. Sun, S.J. Mumby, J.R. Maple, A.T. Hagler, An ab initio CFF93 all-atom force field for polycarbonates, Journal of the American Chemical Society, 116 (1994) 2978–2987.

[26] H. Sun, Ab initio calculations and force field development for computer simulation of polysilanes, Macromolecules, 28 (1995) 701–712.

[27] H. Sun, P. Ren, J. Fried, The COMPASS force field: parameterization and validation for phosphazenes, Computational and Theoretical Polymer Science, 8 (1998) 229–246.

[28] H. Sun, Z. Jin, C. Yang, R.L. Akkermans, S.H. Robertson, N.A. Spenley, S. Miller, S.M. Todd, COMPASS II: extended coverage for polymer and drug-like molecule databases, Journal of Molecular Modeling, 22 (2016) 47.

[29] H. Sun, COMPASS: an ab initio force-field optimized for condensed-phase applications—overview with details on alkane and benzene compounds, The Journal of Physical Chemistry B, 102 (1998) 7338–7364.

[30] D. Rigby, H. Sun, B. Eichinger, Computer simulations of poly(ethylene oxide): force field, PVT diagram and cyclization behaviour, Polymer International, 44 (1997) 311–330.

[31] W.C. Swope, H.C. Andersen, P.H. Berens, K.R. Wilson, A computer simulation method for the calculation of equilibrium constants for the formation of physical clusters of molecules: Application to small water clusters, The Journal of Chemical Physics, 76 (1982) 637–649.

[32] D. Frenkel, B. Smit, Understanding Molecular Simulation: From Algorithms to Applications, Academic Press, San Diego, 1996.

[33] S. Nosé, M. Klein, Constant pressure molecular dynamics for molecular systems, Molecular Physics, 50 (1983) 1055–1076.

[34] S. Nosé, A molecular dynamics method for simulations in the canonical ensemble, Molecular Physics, 52 (1984) 255–268.

[35] S. Nosé, A unified formulation of the constant temperature molecular dynamics methods, The Journal of Chemical Physics, 81 (1984) 511–519.

[36] W.G. Hoover, A.J. Ladd, B. Moran, High-strain-rate plastic flow studied via nonequilibrium molecular dynamics, Physical Review Letters, 48 (1982) 1818.

[37] W.G. Hoover, Canonical dynamics: Equilibrium phase-space distributions, Physical Review A, 31 (1985) 1695.

[38] H.J.C. Berendsen, J.P.M. Postma, W.F. van Gunsteren, A. DiNola, J.R. Haak, Molecular dynamics with coupling to an external bath, The Journal of Chemical Physics, 81 (1984) 3684–3690.

[39] M.R. Hestenes, E. Stiefel, Methods of conjugate gradients for solving linear systems, Journal of Research of the National Bureau of Standards, 49 (1952) 409–436.

[40] P.P. Ewald, Die Berechnung optischer und elektrostatischer Gitterpotentiale, Annalen der Physik, 369 (1921) 253–287.

[41] R.H. Boyd, An off-lattice constant-pressure simulation of liquid polymethylene, Macromolecules, 22 (1989) 2477–2481.

[42] M.A. Aegerter, N. Leventis, M.M. Koebel, Aerogels handbook, Springer Science & Business Media, 2011.

[43] M.P. Allen, D.J. Tildesley, Computer Simulation of Liquids, 2nd ed., Oxford University Press, Oxford, 2017.

[44] L.W. Hrubesh, R.W. Pekala, Thermal properties of organic and inorganic aerogels, Journal of Materials Research, 9 (1994) 731–738.

[45] S. Wang, Y. Xu, M. Yan, Z. Zhai, B. Ren, L. Zhang, Z. Liu, Comparative study of metal-doped carbon aerogel: Physical properties and electrochemical performance, Journal of Electroanalytical Chemistry, (2017).

[46] J. Montes, F. Cuevas, J. Cintas, P. Urban, Electrical conductivity of metal powders under pressure, Applied Physics A, 105 (2011) 935–947.

[47] N. Job, R. Pirard, J. Marien, J.-P. Pirard, Synthesis of transition metal-doped carbon xerogels by solubilization of metal salts in resorcinol–formaldehyde aqueous solution, Carbon, 42 (2004) 3217–3227.

[48] A.T. García-Sosa, M. Castro, Density functional study of $FeO_2$, $FeO_2^+$, and $FeO_2^-$, International Journal of Quantum Chemistry, 80 (2000) 307–319.

[49] A.V. Arbuznikov, M. Hendrickx, L.G. Vanquickenborne, Quantum chemical study of the geometric and electronic structure of the $FeC_2$ molecule, Chemical Physics Letters, 310 (1999) 515–522.

# An All-Atom Model of the Resorcinol–Formaldehyde Aerogel Skeleton from DFT and Molecular Dynamics: Validation and Iron Incorporation

Shoeb ATHAR[1,2], Jérémy GUAZZAGALOPPA[1,2], and Philippe JUND[1,*]

[1] ICGM, Univ Montpellier, CNRS, ENSCM, 34293 Montpellier, France

[2] HUTCHINSON SA - Rue Gustave Nourry, 45120 Châlette-sur-Loing, France

* Corresponding author. E-mail address: philippe.jund@umontpellier.fr (Philippe JUND)

# Supplementary Information

*Contents: S1. Computational details; S2.* Energy-conservation test for the choice of the time step

.

## S1. Computational Details

### S1.1 Force fields compared

The four force fields compared in this study are described below.

The Dreiding force field is a purely diagonal force field with harmonic valence terms and a torsion term. The van der Waals interactions are described by the Lennard-Jones potential, the electrostatic interactions by a Coulomb term, and the hydrogen bond by an explicit hydrogen-bond term [S1].

The polymer consistent force field (PCFF) was developed on the basis of an older force field, CFF91, and is intended for studies of polymers and organic materials. It is useful for polycarbonates, melamine resins and polysaccharides, as well as for carbohydrates, lipids and nucleic acids, and can also be used for cohesive energies, mechanical properties, compressibilities, heat capacities and elastic constants [S2, S3].

The consistent-valence force field (CVFF) is a generalized valence force field whose parameters are suited to amino acids, water and various other functional groups. It is also designed for gas-phase structures and handles peptides, proteins and a wide range of organic systems. It is mainly intended for studies of structures and binding energies, although it is also effective for the calculation of vibrational frequencies and conformational energies [S4].

The Condensed-phase Optimized Molecular Potentials for Atomistic Simulation Studies (COMPASS) force field is the first *ab initio* force field that enables an accurate and simultaneous prediction of gas-phase properties (structural, conformational, vibrational, *etc.*) and condensed-phase properties (equation of state, cohesive energies, *etc.*) for a wide range of molecules and polymers. Its parameterization is carried out in two phases. In the first, *ab initio* phase, the atomic partial charges are calculated from *ab initio* electrostatic potentials and the valence parameters are determined from the total energies and their first and second derivatives with respect to the atomic coordinates, the Lennard-Jones and other van der Waals parameters being set to initial values taken from the force field. In the second phase, the force field is optimized to obtain good agreement with experimental data; for example, the van der Waals parameters are optimized to match the properties of the condensed phase, on the basis of molecular dynamics simulations of liquids for covalent molecular systems and of the energy minimization of crystals for inorganic systems. COMPASS II is a significant extension of the COMPASS force field and includes many more compounds of interest for polymers, porous materials and ionic liquids [S5–S8].

### S1.2 Pair and radial distribution functions

The pair distribution function $g(r)$ describes how the atomic density varies with the distance $r$ from a reference atom [S9]:

$$g(r) = \frac{dn}{dn_0} = \frac{1}{\rho_0\, 4\pi r^2}\ \frac{1}{N}\left\langle \sum_i \sum_{i\prime \neq i} \delta\left(r - \| \mathbf{r}_{i\prime} - \mathbf{r}_i \|\right) \right\rangle = \frac{\rho(r)}{\rho_0} \tag{S1}$$

where $\rho(r)$ is the local atomic density at a distance $r$ from an atom and $\rho_0$ the mean atomic density. $g(r)$ can also be measured experimentally by diffraction.

The radial distribution function, $\mathrm{RDF}(r)$, gives the number of atoms per unit distance around a reference atom and therefore gives access to coordination numbers:

$$\mathrm{RDF}(r) = \frac{dn}{dr} = \frac{1}{N}\left\langle \sum_i \sum_{i\prime \neq i} \delta\left(r - \| \mathbf{r}_{i\prime} - \mathbf{r}_i \|\right) \right\rangle = 4\pi r^2 \rho_0\, g(r) \tag{S2}$$

The mean number of atoms located in a spherical shell centred on a given atom, with inner radius $r_i$ and outer radius $r_e$, is then:

$$N_c = \int_{r_i}^{r_e} \mathrm{RDF}\,(r)\, dr \tag{S3}$$

For a polyatomic system, the partial function involving only atoms of types $\alpha$ and $\beta$, with atomic concentrations $c_\alpha$ and $c_\beta$, is written:

$$\mathrm{RDF}_{\alpha\to\beta}(r) = 4\pi r^2\rho_0\, c_\beta\, g_{\alpha\beta}(r) = \frac{1}{Nc_\alpha}\left\langle \sum_{i_\alpha}\sum_{i_\beta\neq i_\alpha}\delta\left(r - \| \mathbf{r}_{i_\alpha} - \mathbf{r}_{i_\beta} \|\right)\right\rangle \tag{S4}$$

In this work, $g(r)$ was used to compare the three DFT structures of the RF base unit (main text, Figure 1), and the Fe–C partial RDF (Eq. S4 with $\alpha$ = Fe and $\beta$ = C) and its integral (Eq. S3) to characterize the environment of the Fe atoms (main text, Figures 14 and 15).

## S2. Energy-conservation test for the choice of the time step

The time step was chosen by following the total energy of the 15,012-atom system in the microcanonical (*NVE*) ensemble over 0.1 ps, for time steps $\Delta t$ from 0.1 to 1.0 fs (Figure S1). For time steps larger than 0.5 fs, the curves do not fluctuate around the same mean value. For time steps up to 0.4 fs they do, whereas the curve for 0.5 fs drifts away from this mean value (Figure S2); the largest time step that conserves the total energy is therefore 0.4 fs.

The root-mean-square fluctuation of the total energy, $\langle \Delta E_{\mathrm{tot}}^2\rangle^{1/2}$, should scale as $\Delta t^2$ for the velocity Verlet algorithm [S9]. This scaling holds for time steps below 0.5 fs but not above (Figure S3). Since the largest time step that conserves the total energy minimizes the calculation time, $\Delta t$ = 0.4 fs was used for all the simulations.

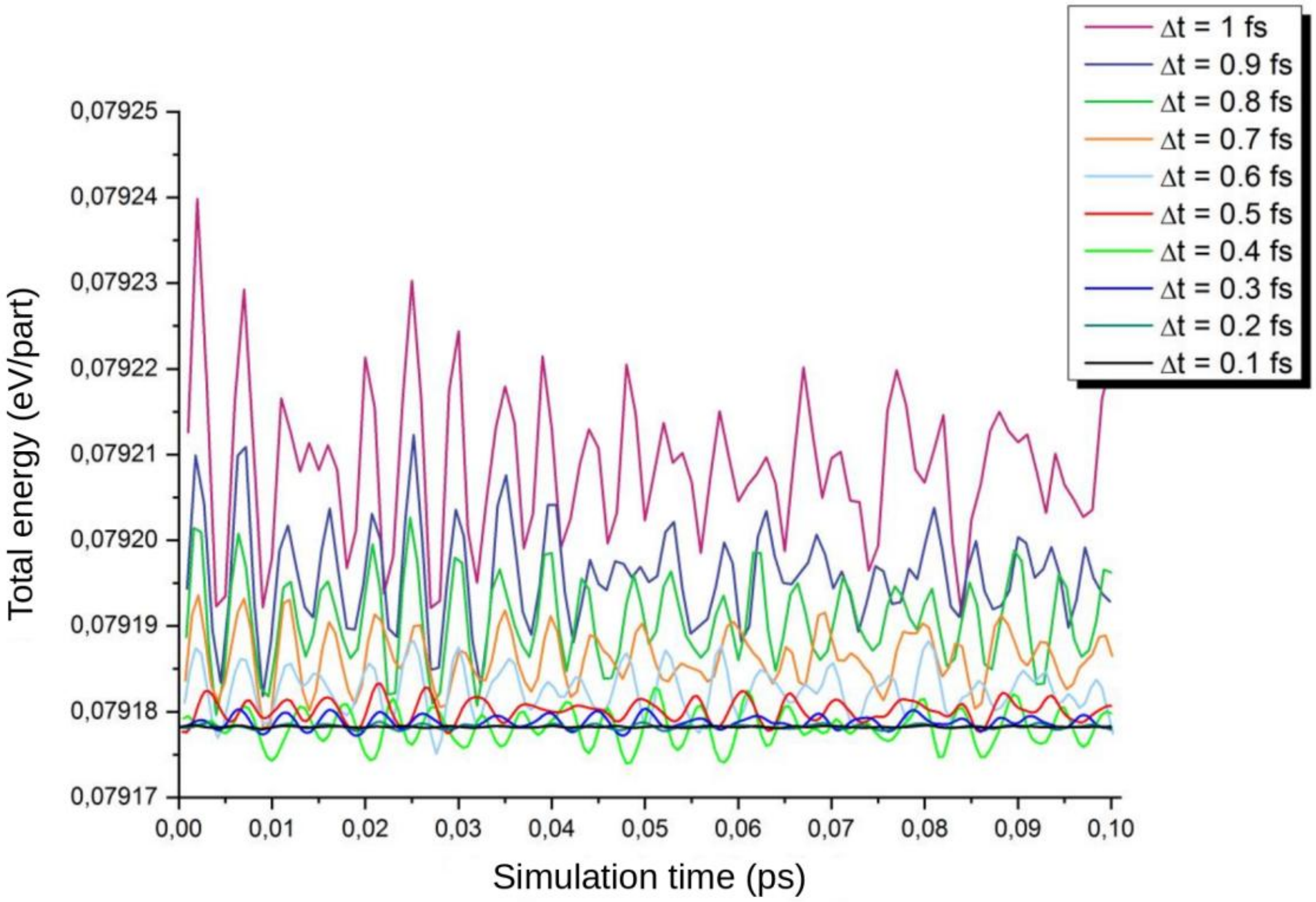

**Figure S1.** Total energy during 0.1 ps *NVE* runs of the 15,012-atom system for time steps $\Delta t$ from 0.1 to 1.0 fs.

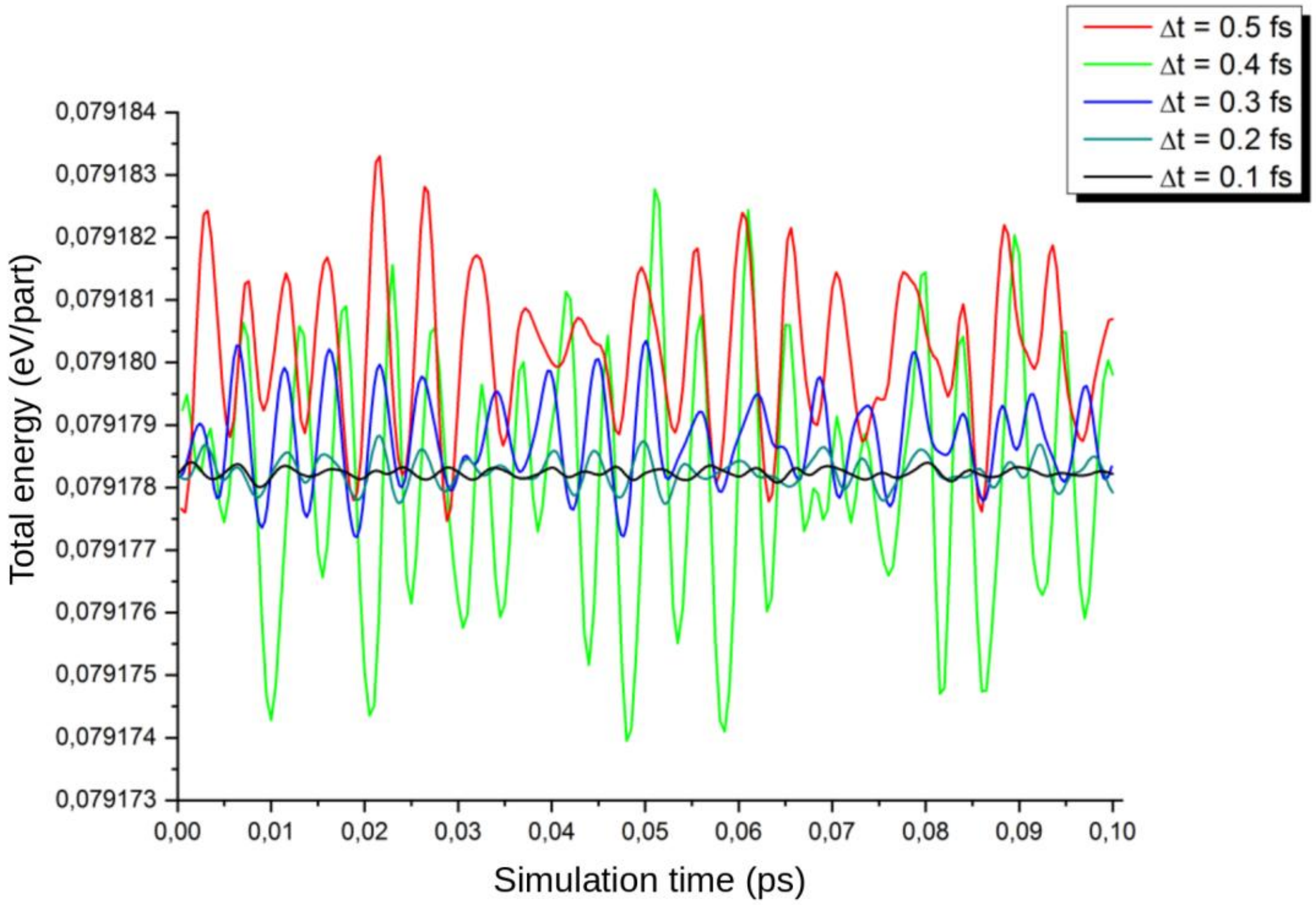

**Figure S2.** Total energy during 0.1 ps *NVE* runs of the 15,012-atom system for time steps $\Delta t$ from 0.1 to 0.5 fs.

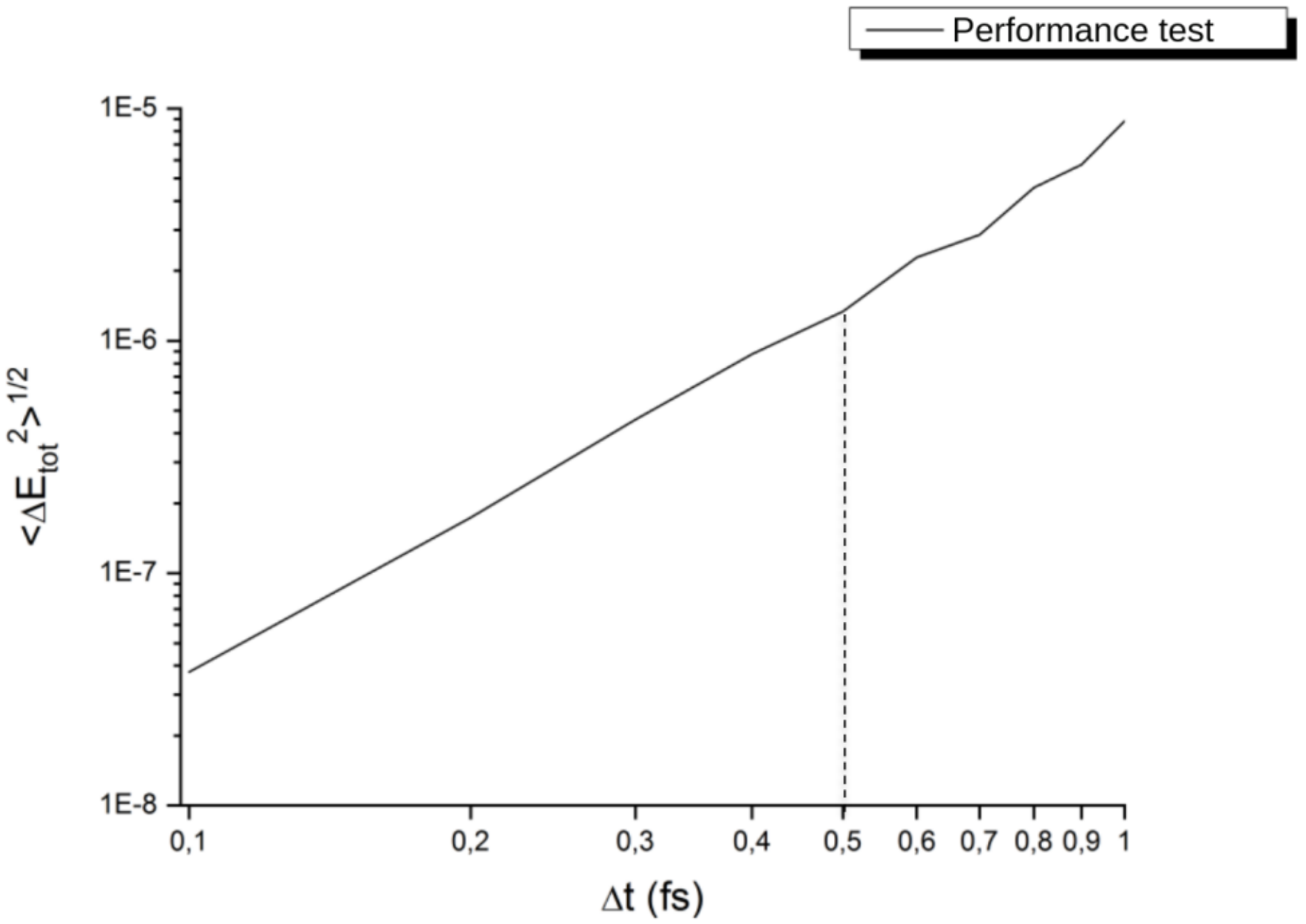

**Figure S3.** Root-mean-square fluctuation of the total energy, $\langle \Delta E_{\mathrm{tot}}^{2} \rangle^{1/2}$, as a function of the time step $\Delta t$.

## References

[S1] S.L. Mayo, B.D. Olafson, W.A. Goddard, DREIDING: a generic force field for molecular simulations, Journal of Physical Chemistry, 94 (1990) 8897–8909.

[S2] H. Sun, S.J. Mumby, J.R. Maple, A.T. Hagler, An ab initio CFF93 all-atom force field for polycarbonates, Journal of the American Chemical Society, 116 (1994) 2978–2987.

[S3] H. Sun, Ab initio calculations and force field development for computer simulation of polysilanes, Macromolecules, 28 (1995) 701–712.

[S4] P. Dauber-Osguthorpe, V.A. Roberts, D.J. Osguthorpe, J. Wolff, M. Genest, A.T. Hagler, Structure and energetics of ligand binding to proteins: Escherichia coli dihydrofolate reductase–trimethoprim, a drug–receptor system, Proteins: Structure, Function, and Bioinformatics, 4 (1988) 31–47.

[S5] H. Sun, P. Ren, J. Fried, The COMPASS force field: parameterization and validation for phosphazenes, Computational and Theoretical Polymer Science, 8 (1998) 229–246.

[S6] H. Sun, Z. Jin, C. Yang, R.L. Akkermans, S.H. Robertson, N.A. Spenley, S. Miller, S.M. Todd, COMPASS II: extended coverage for polymer and drug-like molecule databases, Journal of Molecular Modeling, 22 (2016) 47.

[S7] H. Sun, COMPASS: an ab initio force-field optimized for condensed-phase applications—overview with details on alkane and benzene compounds, The Journal of Physical Chemistry B, 102 (1998) 7338–7364.

[S8] D. Rigby, H. Sun, B. Eichinger, Computer simulations of poly(ethylene oxide): force field, PVT diagram and cyclization behaviour, Polymer International, 44 (1997) 311–330.

[S9] M.P. Allen, D.J. Tildesley, Computer Simulation of Liquids, 2nd ed., Oxford University Press, Oxford, 2017.